\documentclass[a4paper,12pt,times,numbered,print,index]{Classes/PhDThesisPSnPDF}

\input{Preamble/preamble}

\title{Automated quantitative analysis of first-pass myocardial perfusion magnetic resonance imaging data}

\author{Cian Michael Scannell}

\dept{School of Biomedical Engineering and Imaging Sciences}

\university{King's College London}
\crest{\includegraphics[width=0.3\textwidth]{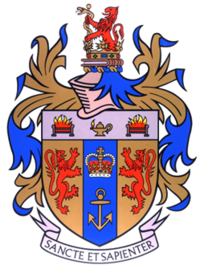}}

\degreetitle{Doctor of Philosophy}

\college{{\textbf{Supervisors:}\newline Dr. A. Chiribiri\newline
Dr. J. Lee}}

\subject{LaTeX} \keywords{{LaTeX} {PhD Thesis} {Engineering} {University of
Cambridge}}

\ifdefineAbstract
\fi

\ifdefineChapter
\fi

\begin{document}

\frontmatter

\maketitle


\begin{declaration}

I hereby declare that except where specific reference is made to the work of 
others, the contents of this dissertation are original and have not been 
submitted in whole or in part for consideration for any other degree or 
qualification in this, or any other university. This dissertation is my own 
work except where explicitly stated otherwise in the text

\end{declaration}

\begin{acknowledgements}      

I am indebted to my primary supervisor Amedeo Chiribiri for his hard-work and kindness through-out the four years of this PhD. None of this would have been possible without his support and guidance. I would also like to thank Jack Lee for his productive comments and advice. 

I am grateful to Philips Healthcare for co-funding my studentship and would like to especially thank Marcel Breeuwer for his mentorship and Torben Schneider for his scanner and software advice.

I would also like to thank all my collaborators for making this work possible. In particular, I was extremely lucky to work with Mitko Veta and Adriana Villa, who taught me so much. Without their help and fruitful discussions, this would have been a much different thesis.

I was fortunate to do this work in the CDT in Medical Imaging and I am grateful for all the extra help that was provided by the CDT. I am also thankful to the close friends that I met here: Pam, Dan, George, and Marina who made this process a lot more enjoyable.

Finally, I would like to thank my family, and most of all my parents, not just for their support during my PhD but for everything leading up to this. I couldn't have done it without you. 

\end{acknowledgements}

\begin{abstract}
Coronary artery disease (CAD) remains the world’s leading cause of mortality and the disease burden is continually expanding, particularly in western countries, as the population ages. Recently, the MR-INFORM randomised trial has demonstrated that the management of patients with stable CAD can be guided by stress perfusion cardiovascular magnetic resonance (CMR) imaging and it is non-inferior to the using the invasive reference standard of fractional flow reserve. The benefits of using stress perfusion CMR include that it is non-invasive and significantly reduces the number of unnecessary coronary revascularisations. As compared to other ischaemia tests, it boasts a high spatial resolution and does not expose the patient to ionising radiation. However, the main limitation of stress perfusion CMR is that the diagnostic accuracy is highly dependent on the level of training of the operator, resulting in the test only being performed routinely in experienced tertiary centres.

The clinical translation of stress perfusion CMR would be greatly aided by a fully-automated, user-independent, quantitative evaluation of myocardial blood flow. This thesis presents major steps towards this goal: robust motion correction, automated image processing, reliable quantitative modelling, and thorough validation. The motion correction scheme makes use of data decomposition techniques, such as robust principal component analysis, to mitigate the difficulties in image registration caused by the dynamic contrast enhancement. The motion corrected image series are input to a processing pipeline which leverages the recent advances in image processing facilitated by deep learning. The pipeline utilises convolutional neural networks to perform a series of computer vision tasks including myocardial segmentation and right ventricular insertion point detection. The tracer-kinetic model parameters are subsequently estimated using a Bayesian inference framework. This incorporates the prior information that neighbouring voxels are likely to have similar kinetic parameters and thus improves the reliability of the estimated parameters. The full process is validated in a well characterised patient population against coronary angiography and invasive measurements. It is shown to be accurate at detecting reductions in myocardial blood flow while further discriminating between patients with no significant CAD and those with obstructed coronary arteries or microvascular dysfunction.
\end{abstract}


\tableofcontents

\listoffigures

\listoftables


\printnomenclature

\mainmatter


\chapter{Introduction}  

\section{Motivation}

Stress perfusion cardiovascular magnetic resonance (CMR) imaging has a class IA indication in the European guidelines for the evaluation of patients with an intermediate risk of coronary artery disease \cite{Montalescot2013a}. However, myocardial perfusion imaging is performed far less frequently with CMR than with the alternative nuclear imaging approaches: single-photon emission computed tomography (SPECT) and positron emission tomography (PET). This is despite the fact that CMR boasts superior spatial resolution, a lack of ionising radiation, and can also provide a full assessment of cardiac function and tissue viability. Furthermore, multiple trials and studies have shown CMR to have at least comparable diagnostic accuracy to SPECT \cite{Greenwood2012,Schwitter2013a} and to be non-inferior to invasive measurements for the management of patients \cite{Nagel2019}.

In fact, stress perfusion CMR is primarily used in highly experienced centres and there is a need to encourage its adoption in less specialised centres. This need, however, is hampered by the dependence of the diagnostic accuracy of the test on the level of training of the operator \cite{Villa2018}. A possible solution to this is the quantification of myocardial perfusion which offers a user-independent assessment of the images that could facilitate the adoption of stress perfusion CMR in less specialised centres.

The aim of this thesis is to develop methods to overcome the technical difficulties of quantitative myocardial perfusion CMR. Challenges such as respiratory motion, the reliability of the analysis, and the time consuming processing have lead to the approach only being used as a research tool and not in clinical practice. This thesis deals with the contrast-enhancement during motion compensation in a principled manner, introduces more robust kinetic parameter estimation approaches and combines this with a deep learning-based image processing pipeline to make the analysis fully automated. The initial clinical validation of the proposed methods is also reported.


\section{Contribution of the thesis}

The original contributions of the thesis can be summarised as follows:

\begin{itemize}
\item \textbf{Robust motion compensation.} The motion compensation of myocardial perfusion CMR, using image registration, is difficult due to the rapidly changing contrast during the passing of the gadolinium bolus. This invalidates the assumptions of the similarity measures optimised in the image registration. This work proposes the use of robust principal component analysis (RPCA) to decompose the baseline signal from the dynamic contrast-enhancement. Image registration is then more readily applied in the absence of the dynamic contrast-enhancement. The proposed approach is evaluated qualitatively by expert clinicians, as well as quantitatively. 
\item \textbf{Fully automated image processing.} Deep learning is used to automate the computer vision tasks required for the quantitative modelling. This includes the automation of the myocardial segmentation, the identification of the arterial input function, and the detection of the right ventricular insertion points. Each step is evaluated individually using a suitable metric and additionally, the full automated pipeline is compared to the manual processing.

\item \textbf{Reliable quantitative modelling.} The identification of the tracer-kinetic parameters from the imaging data is an ill-posed inverse problem, and as such, the estimated values are subject to uncertainties \cite{Buckley2002}. This work introduces Bayesian inference as a principled approach to incorporate prior information in the parameter estimation. In particular, the prior knowledge that neighbouring pixels are likely to have similar kinetics is used to constrain the parameter estimation and improve reliability.

\end{itemize}

Furthermore, an initial assessment of the performance of the  proposed pipeline in comparison to invasive coronary angiography. This is performed on an external validation set (i.e, none of the patient data considered in this assessment has been used in the development of any of the underlying methods) to give an accurate reflection of the clinical performance.  

\newpage
\section{Outline of the thesis}
The thesis is organised into 7 further chapters, the contents of which are:

\begin{itemize}
\item \textbf{Chapter 2:} a brief background on the physiology and pathophysiology of coronary artery disease is first provided. The chapter subsequently gives an overview of some of the imaging approaches to diagnosing coronary artery disease.
\item \textbf{Chapter 3:} gives an introduction to myocardial perfusion CMR. This includes the background on the MR acquisition of the data, the tracer-kinetic modelling, and a review of the state-of-the-art approaches in the literature. 
\item \textbf{Chapter 4:} introduces the problem of motion compensation of myocardial perfusion CMR, and describes RPCA and how it is used. The proposed approach to the problem is then described and evaluated.
\item \textbf{Chapter 5:} firstly, introduces the image processing problems and gives the requisite deep learning background. It then describes the automated image processing pipeline and considers its application and evaluation. The pipeline is further compared to the manual processing by expert operators.
\item \textbf{Chapter 6:} discusses the limitations of the conventional parameter estimation methods. Then, the Bayesian approach with spatial priors is developed and evaluated both in simulated and patient data.
\item \textbf{Chapter 7:} applies the methods developed in Chapters 4-6 prospectively to an initial patient cohort and reports the diagnostic accuracy in comparison to the invasive measurements.
\item \textbf{Chapter 8:} summarises the contributions made in this thesis and considers both the future work required and the potential clinical impact of that work.
\end{itemize}

\chapter{Background}

\section[Cardiac anatomy and function]{Cardiac anatomy and function}

\begin{figure}[htbp!] 
\centering    
\includegraphics[width=0.75\textwidth]{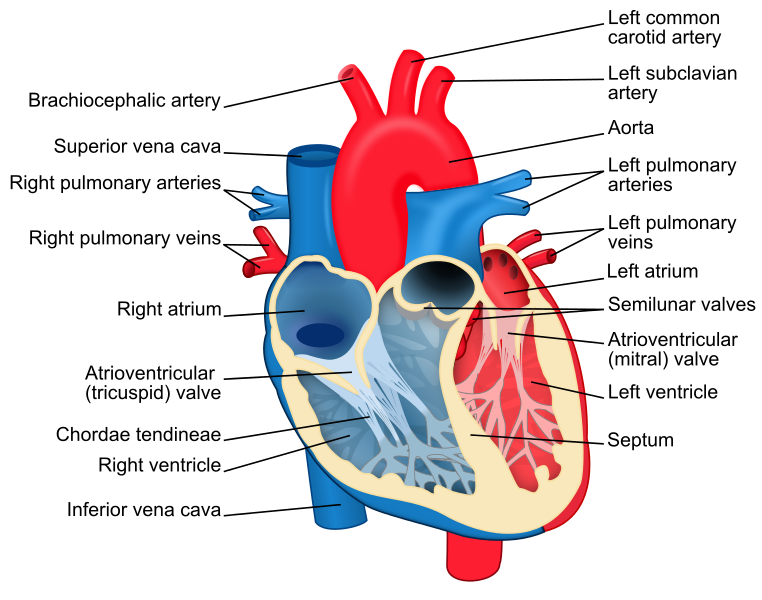}
\caption[The anatomy of the heart]{The anatomy of the heart with the chambers, major vessels and valves labelled. The colour blue indicates the structure is a constituent part of the pulmonary circulatory system and similarly red identifies the systemic circulatory system, taken from \cite{WikiHeart}.}
\label{fig:heart_anatomy}
\end{figure}

The heart is the centre of the cardiovascular system. It acts as a pump in order to distribute oxygen and nutrients to the tissue around the body and to subsequently remove waste products \cite{Levick2010}. This function is essential for sustaining activity in the tissue and to avoid tissue necrosis.  The heart is responsible for the circulation of both oxygenated and deoxygenated blood and does so using a series of chambers and vessels, as shown in Figure~\ref{fig:heart_anatomy}. In order to deal with both the distribution of oxygenated blood and the reception of deoxygenated blood, the heart is divided into two systems: the pulmonary circulatory system and the systemic circulatory system. The pulmonary circulation starts at the right atrium from which the right ventricle is filled with deoxygenated blood. The right ventricle then pumps this blood to the lungs through the pulmonary arteries. The blood is oxygenated in the lungs and returns to the heart, specifically the left atrium, through the pulmonary veins. In the systemic circulation, the left ventricle is filled from the left atrium and the blood is pumped, through the aorta, around the body. Deoxygenated blood returns to the heart through the superior and inferior venae cavae to complete the cycle.

\begin{figure}[htbp!] 
	\centering    
	\includegraphics[width=0.6\textwidth]{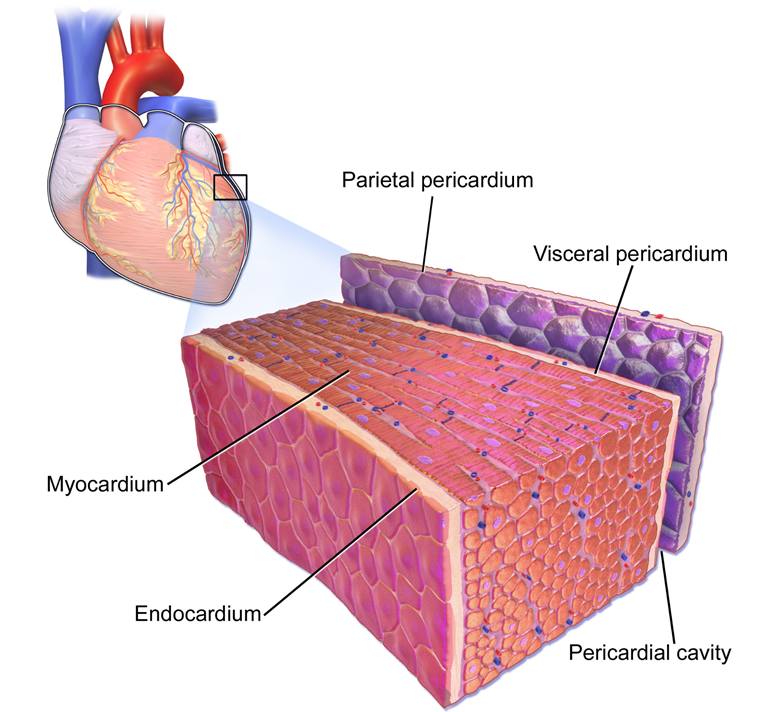}
	\caption[The cardiac muscle]{The underlying structure of the cardiac muscle, adapted from \cite{WikiHeartMuscle}.}
	\label{fig:heart_muscle}
\end{figure}

The heart muscle, as shown in Figure~\ref{fig:heart_muscle}, is made up of three layers: a thick middle layer known as the myocardium, the inner endocardium, and the outer epicardium (also known as the visceral pericardium). As with all other tissue, cardiac muscle requires oxygen and nutrients in order to sustain viability and keep pumping blood. The cardiac tissue does not receive oxygen and nutrients by simply holding oxygenated blood. Thus, the requisite oxygen and nutrients need to be delivered to the muscle and this is done through the coronary circulatory system.

\section[The coronary circulatory system]{The coronary circulatory system} \label{sec:coronaries}
The passage of blood between the chambers of the heart is controlled by four valves, shown in Figure~\ref{fig:heart_anatomy}, which separate the atria from the ventricles and the ventricles from a blood vessel. The valves are made of leaflets (flaps) which are opened and closed by changing flow and pressure. The left ventricle is separated from the aorta by the aortic valve. On the leaflets of the aortic valve lies the coronary ostium, the openings of the left and right coronary arteries. The left coronary artery (LCA) branches into the left anterior descending (LAD) and left circumflex (LCx) branches. The right coronary artery (RCA), the LAD, and LCx further subdivide in order to cover the breadth of the epicardium and distribute oxygen and nutrients to the muscle. Though the exact anatomical structure varies from person to person, generally, the RCA supplies the right atrium, right ventricle, and the back of the septum. The front of the septum is supplied by the LAD, as well as the front and bottom of the left ventricle. The left atrium and the back and side of the left ventricle is supplied by the LCx \cite{Feigl1983}. 

The coronary circulatory system performs a particularly important role as the myocardium demands 20 times more oxygen than skeletal muscle, a peak heart rate \cite{Levick2010}. To satisfy this high demand, even under normal resting conditions the myocardium needs to extract 70-80\% of the available oxygen \cite{CPC_CorAnatBF}. During exercise, the rate at which the heart beats increases in order to satisfy the increased systemic demand for oxygen and thus the myocardium also demands more oxygen. Since the extraction rate of oxygen from blood is already high, there is limited scope for increased extraction and the increased demand must be met by increasing coronary blood flow \cite{Levick2010}.

\begin{figure}[htbp!] 
	\centering    
	\includegraphics[width=0.65\textwidth]{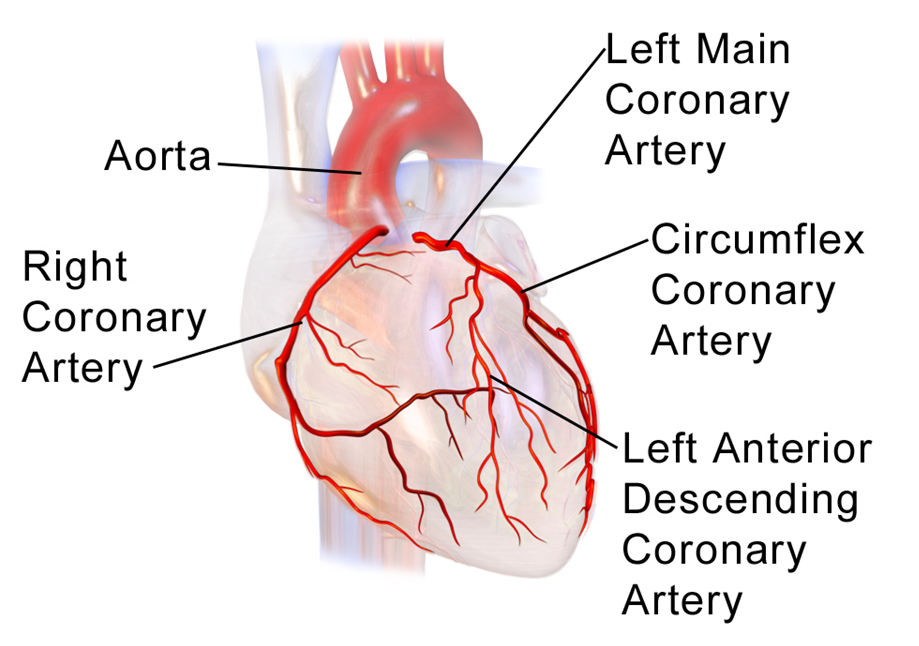}
	\caption[The coronary circulation system]{The coronary arteries shown traversing the wall of the heart, adapted from \cite{WikiCoronaries}.}
	\label{fig:coronaries}
\end{figure}

Coronary blood flow is primarily driven by the pressure gradient between the aorta and the right atrium \cite{Feigl1983}. Since the coronaries are constricted when the is heart contracted (systole) most flow occurs when the heart is relaxed (diastole). This further inhibits coronary blood flow at higher heart rate. As the heart rate increases, the diastolic proportion of the cardiac cycle falls faster than systole, leading to less unrestricted flow \cite{Levick2010}. Blood also flows faster through wider vessels, as there is less resistance. Natural or pharmacologically-induced vasodilation thus increase blood flow \cite{Feigl1983}. Conversely, the narrowing of the vessels can lead to reduced flow. The coronary arteries can be narrowed by atherosclerosis. This is a disease in which there is a build-up of plaque in the walls of the vessels.

The heart has an innate ability to adapt to changing circumstances. In a processes known as autoregulation, blood flow is well regulated in spite of changes in the available pressure gradient \cite{CPC_Auto}. For example, in the case of the narrowing of a vessel, there will be a proportional increase in coronary perfusion in order to maintain roughly constant flow. From a myogenic point of view, the vessel stretching caused by the increased pressure leads to the depolarisation of the cells and subsequent vasoldilation \cite{CPC_Auto}. Metabolic mechanisms of autoregulation are also at play and stem from the coupling between metabolic activity and blood flow. Hypoxia, a oxygen deficiency, is known to cause vasodilation as it leads to the release of adenosine, which relaxes the vessels. Hypoxia is also thought to open adenosine triphosphate-sensitive potassium (K$^+$) channels and the increase in interstitial K$^+$ dilates the vessels \cite{CPC_Vaso}.

\section[Ischaemic heart disease]{Ischaemic heart disease}

Myocardial ischaemia is the process ensuing when coronary blood flow fails to meet the muscles oxygen and nutrient requirements. The tissue becomes hypoxic and if sustained can undergo irreversible cell necrosis, known as a myocardial infarction or more commonly, a heart attack. 

As discussed, ischaemia does not occur under ideal, healthy conditions due to the elegant coupling between oxygen requirements and coronary blood flow. However, such ideal conditions are not guaranteed and in particular are precluded by the atherosclerosis of the coronary arteries. The cholesterol-rich plaque, atheroma, builds up naturally over time in the vessels, though it can be significantly accelerated by lifestyle choices. This build-up can lead to a reduction in blood flow known as ischaemic heart disease (IHD) or coronary artery disease (CAD). Cardiovascular disease (CVD), of which CAD is the most prevalent, is the leading cause of death globally. It accounted for an estimated 17.9 million deaths in 2016 and there is expected to be more hospitalisations and deaths as the population ages, particularly in western countries \cite{WHO_CVD}.

The increased pressure caused by the lumen narrowing can be well regulated under resting conditions. It could be that even for moderately narrowed vessels, the increased resistance in the vessel can be offset by distal vasodilation \cite{Levick2010}. However, for significantly narrowed vessels, coronary blood flow or perfusion (which are used synonymously in this thesis) will be reduced. The problem is exacerbated during stress or exercise as increased flow is required but the distal vasodilation is not sufficient to achieve this. The effect of a single stenosis on the myocardium can be particularly pronounced due to the limited flow between branches of the coronary tree \cite{Levick2010}. The result being that most of the tissue downstream of a stenosis will be affected.

Angina (pectoris) is the term used to describe the chest pain associated with myocardial ischaemia and is in nature either stable or unstable. Stable angina is typically absent at rest and increases with increasing stress or physical exertion. Unstable angina is less predictable. It can be caused by the built-up plaque rupturing and forming blood clots which cut off flow \cite{Detry1996}. This may be unrelated to physical exertion and there may have been no prior warning. 

\begin{figure}[htbp!]
	\centering
	\includegraphics[trim=1cm 9cm 1cm 9cm,width=.7\textwidth]{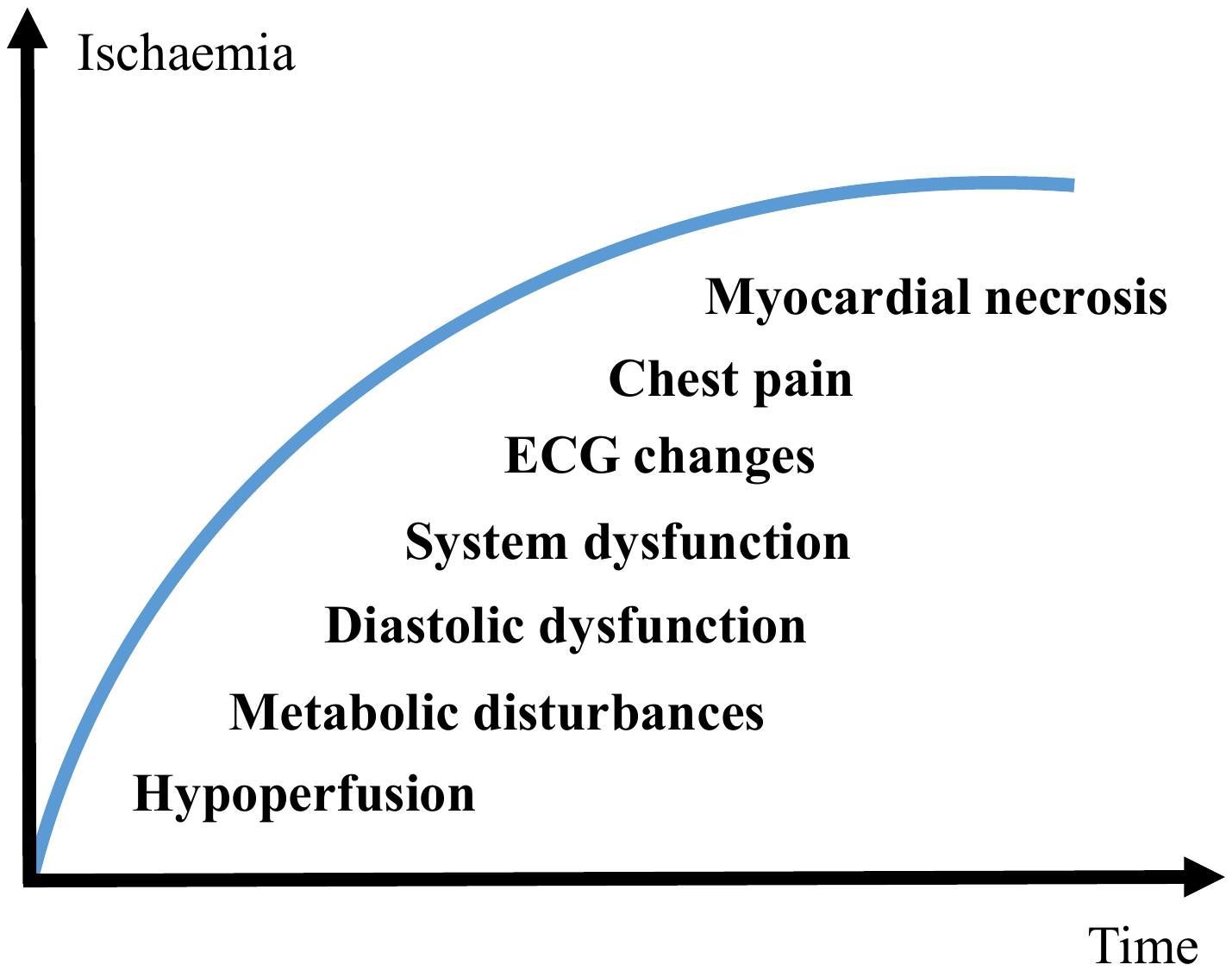}
	\caption[The ischaemic cascade]{A schematic representation of the  ischaemic cascade}
	\label{fig:ic}
\end{figure}

The damage associated with myocardial ischaemia can be reversed if the oxygen deprivation is transient. However, if sustained for a period of time, the ischaemia will lead to a myocardial infarction. The oxygen deprivation does not instantaneously lead to cell necrosis, rather it triggers a series of predictable events, commonly referred to as the ischaemic cascade. The series of events, shown in Figure~\ref{fig:ic}, begins with reduced perfusion which will in turn lead to impaired diastolic function, (subsequently) impaired systolic function, and culminate in angina and myocardial infarction.

The ischaemic cascade is central to any discussion on the diagnosis of CAD as earlier diagnosis (and treatment) is known to improve outcomes \cite{Ansari2009}. While traditional clinical assessment is concerned with angina and ECG changes, as will be discussed in this thesis, state-of-the-art imaging techniques are focussed on identifying earlier and earlier markers of CAD.

Myocardial infarctions are classified based on their effect on the electrocardiogram (ECG) of the patient. An elevation of the ST segment of the ECG can be caused by a thrombus, a completely blocked lumen leading to transmural ischaemia and a so-called ST-elevation myocardial infarction (STEMI) \cite{CPC_EP}. Non-transmural ischaemia can lead to a non ST-elevation myocardial infarction (NSTEMI), this is typically less dangerous as it affects less of the myocardium.

The result of an infarction is that the tissue forms a fibrotic scar \cite{CPC_MI}. This non-viable tissue has a reduced ability to conduct electrical activity and results in the ventricles being unable to contract properly. The impaired ventricular function causes a reduction in cardiac output and can lead to death \cite{Detry1996}.

\section[Imaging coronary artery disease]{Imaging coronary artery disease}
Imaging for the diagnosis of CAD follows two main directions: anatomical imaging and functional imaging. Anatomical imaging is focused on the visualisation of the coronary arteries and the narrowing thereof. Fnctional imaging assess the functional significance of the narrowing arterial lumen. Functional tests are usually performed during stress in order to induce perfusion or wall motion abnormalities. 
The ideal test would, as well as being both sensitive and specific, be safe. With this in mind, non-invasive imaging is preferable as it reduces the risk of adverse advents as a result of the testing and the test ideally would limit the patients exposure to ionising radiation. In order to maximise the sensitivity of the test it would be desirable for the test to probe effects that are observable at early stages of the ischaemic cascade. In particular, with regards to the functional testing, the ability to detect subtle effects at earlier stages of disease is advantageous and requires either high spatial or temporal resolution.

The history of CAD imaging began with the use of anatomical imaging and then progressed towards the use of functional imaging to diagnose CAD earlier. However, there has been a recent move back towards anatomical imaging, especially in order to rule out CAD. This is based on the argument that even after an abnormal functional test, some form of anatomical imaging is required to confirm the diagnosis of CAD. However, as discussed, this does not provide information on the haemodynamical consequences of the disease and indeed the future probably lies in a combined functional and anatomical assessment.

The following subsections will detail the available forms of CAD testing and include a brief discussion of their respective pros and cons.

\subsection{X-ray coronary angiography}
The coronary angiography is considered to be the gold standard for the diagnosis of CAD \cite{Fihn2012}. Catheters are inserted into the patients arteries through either the radial or femoral artery and guided to the ascending aorta. An iodine-based radio-opaque contrast agent is then injected in order to visualise the blood vessels and potential narrowings on the X-ray images. While the test is highly diagnostic, the major downsides of the procedure are the risk of complications, including puncturing a vessel, the use of ionising radiation, and the discomfort caused to the patient. 

There are further benefits to the approach. It is possible to derive quantitative measures such as the fractional flow reserve (FFR). For this, a pressure sensor is used on the tip of the wire inserted and the FFR value is computed as the ratio of the pressures measured in the aorta and immediately downstream of a lesion. Furthermore, upon the identification of a haemodynamically significant lesion, it is possible to treat it with a stent or a balloon in a procedure known as percutaneous coronary intervention (PCI).

\subsection{Electrocardiogram}
The ECG monitors the electrical activity in and around the heart. As previously discussed, myocardial infarctions can alter the way that the heart conducts electrical activity and this can be observed on an ECG. Since the test is cheap and safe, it is often one of the first assessments that a patient receives \cite{Braunwald2015}. The limitation of the ECG is that the changes in electrical activity only occur at an advanced stage of the ischaemic cascade, though ischaemia can be detected under exercise stress. 

\subsection{Echocardiogram}
The echocardiogram (echo) is a cardiac ultrasound and uses a transducer to send and receive ultrasound beams to allow the visualisation of cardiac anatomy. It provides a quick and easy assessment of cardiac structures, wall motion, and, using Doppler-based imaging, even blood flow \cite{Votavova2015}. An echo is cheap, free from ionsing radiation, and the devices can be brought to a patients bedside. As a result, it is the most commonly performed imaging assessment of the heart \cite{Braunwald2015}. The main limitations to the general applicability of echocardiography in clinical practice are that the image quality depends on the anatomy and acoustic window of the patient, the positioning of the transducers, and the skill of the operator. The images can suffer from a significant amount of speckle noise and may, thus, not be of diagnostic quality. Furthermore, the wall motion abnormalities probed by the test also only manifest themselves at a late stage of the ischaemic cascade.

\subsection{Single-photon emission computed tomography}
SPECT is one of the most commonly used cardiac imaging techniques. It is a particularly important imaging modality to consider as it can assess myocardial blood flow \cite{Beller2011}. This is crucial for the early diagnosis, and to guide the management, of patients with CAD as impaired perfusion occurs early in the ischaemic cascade. SPECT counts the number of photons emitted in an area (and thus the amount of radioisotope accumulated), this is assumed to be proportional to blood flow in the tissue. The imaging is typically carried out with the use of vasodilator stressor agents in order to identify stress-inducible ischaemia.

The major limitation of the modality is that the spatial resolution achievable is typically on the order of 10mm$^3$. This makes it difficult to identify regional perfusion abnormalities and the images are subject to motion and partial volume effects. Additionally, long imaging times required to record sufficient signal and the photon emitting radioisotopes injected typically have long half-lives meaning significant exposure to ionising radiation for the patient \cite{Beller2011}.

\subsection{Positron emission tomography}
PET is, in principle, similar to SPECT except that the radiotracers are labelled with positron emitting isotopes. The signal recorded, by an array of detectors around the body, is from the gamma rays emitted when the positrons collide with the electrons in the tissue \cite{Braunwald2015}. Metabolism can be quantified in absolute terms and it is also possible to derive quantitative values of myocardial blood flow (MBF) in units of millilitres per minute per gram of tissue (ml/min/g) \cite{Schuijf2005}. The quantitative values easily identify areas of ischaemia while the patient is stressed. The array of detectors yield superior spatial resolution to SPECT and furthermore, the short half-lives of the radiotracers used, as compared to SPECT, means less exposure to ionising radiation for the patient. However, the amount of radiation and spatial resolution are still far from ideal.

\subsection{Computed tomography}
Computed tomography (CT) has developed into an emerging technology in the application of non-invasive CAD diagnosis. In particular, coronary CT angiography (CCTA) facilitates the visualisation of the coronary arteries and the National Institute for Health and Care Excellences (NICE) guidelines also include fractional flow reserve derived from CCTA (FFRct). FFRct can determine the functional significance of a lesion \cite{Moss2017}. Recent improvements in CT detector rows have made it possible to test for ischaemia using dynamic stress perfusion CT \cite{George2009}. However, this is yet to see widespread adoption due to concerns about the radiation dose, technical difficulties leading to motion artefacts, a low contrast-to-noise ratio, quantification challenges, and a lack of availability \cite{Dewey2020}.

\subsection{Cardiovascular magnetic resonance}
Magnetic resonance imaging (MRI) is a hugely versatile imaging modality and CMR has the potential to overcome many of the limitations of the imaging modalities discussed so far. In particular, myocardial perfusion MRI, the subject of this thesis, has emerged as a sensitive and specific ischaemia test. The benefits of myocardial perfusion MRI over the other ischaemia tests discussed are that it is free from ionising radiation and gives high spatial resolution. Perfusion is typically assessed with dynamic contrast-enhanced imaging, using a Gadolinium-based contrast agent. 

The MR-IMPACT II trial showed perfusion MRI to have a similar diagnostic accuracy to SPECT in a multi-centre setting \cite{Schwitter2013a} and similarly the CE-MARC trial found multi-parametric CMR to be more accurate than SPECT for the diagnosis of coronary artery disease \cite{Greenwood2012}. While the CE-MARC 2 trial showed CMR imaging is associated with a lower probability of unnecessary coronary angiography than SPECT imaging without an increase in major adverse cardiac events (MACE) \cite{Greenwood2016}. It further found that functional testing outperformed anatomical tests with respect to the endpoint of unnecessary angiography. Recently, the MR-INFORM (MR perfusion imaging to guide management of patients with stable coronary disease) study randomised 918 patients with typical angina to either FFR-guided management or perfusion CMR-guided management \cite{Nagel2019}. It found perfusion CMR to be non-inferior to FFR for the management of patients with respect to major adverse cardiac events. Additionally, it found that a perfusion MRI-guided approach significantly reduced the number of unnecessary coronary revascularisations. Further adding to the evidence supporting the use of perfusion CMR, the SPINS (stress CMR perfusion imaging in the United States) study retrospectively analysed data from the Society for Cardiovascular Magnetic Resonance (SCMR) registry. In this cohort of nearly 2500 patients across 13 centres, the SPINS study showed the long-term prognostic performance of perfusion MRI in a real-world setting \cite{Kwong2019}. It also showed the costs benefits of CMR as a gate-keeper test: very low downstream costs for ischaemia testing were reported for patients after negative CMR tests.

Further to the evidence presented, the unique selling point of perfusion CMR may be that it can be incorporated easily into a comprehensive CMR examination. The modality's wide-ranging utility has led to it being described a "one-stop shop" for the assessment of ischaemic heart disease \cite{Kramer1998}. In a single examination, it is possible to image ventricular function, myocardial ischaemia using stress perfusion imaging and myocardial viability using late gadolinium enhancement. In the coming years, further technical development will also see CMR be used more frequently for the visualisation of the coronary arteries \cite{Bustin2020}.

Despite all the apparent benefits of stress perfusion CMR, it is still not common place in clinical practice. Thus far, all evidence has been accumulated at expert tertiary centres where there is a vast experience of performing and reporting the scans. One of the reasons limiting the widespread clinical adoption is that there is limited data supporting its use at less specialised centres. The reading of the scans is complex and time consuming, and there is little access to training. A recent study by Villa et al. \cite{Villa2018} showed the diagnostic accuracy is highly dependent on the level of training of the reader.

This serves to highlight the need for an objective, user-independent assessment of ischaemia and in particular, the quantitative analysis of stress perfusion CMR. Quantitative measures of myocardial perfusion have the potential to add the benefits of speed, automation, and reproducibility to a test which is already known to be accurate, non-invasive, and free from ionising radiation. The quantitative values also add independent prognostic value \cite{Sammut2017,Knott2020}. However, quantitative perfusion CMR is still hindered by technical difficulties such as respiratory motion, the time-consuming nature of the image processing, and questions about the reliability of the quantitative values. The solutions to these challenges will be discussed in detail in this thesis.
\chapter{Myocardial perfusion MR imaging}
The chapter will introduce the basic concepts of MRI and how these relate to the imaging of myocardial perfusion. It will then discuss the 
theory underlying the quantification of myocardial perfusion: the tracer-kinetic modelling, parameter inference, and arterial input function estimation. The chapter concludes with a review of the state-of-the-art in quantitative myocardial perfusion.

\section{The basics of MRI}
MRI is based on the principle of nuclear magnetic resonance (NMR). NMR is the effect exerted on nucleons with non-zero angular spin when exposed to a magnetic field. Atoms with an odd mass number (an atoms mass number is the total number of protons and neutrons in its nucleus) have half-integer spin and this spin is aligned in the presence of an external magnetic field $\mathbf{B_0}$. This yields a net change in magnetic moment $\mathbf{M} = [M_x, M_y, M_z]$ described by:
\begin{align}
	\diff{\mathbf{M}}{t} =  \gamma \mathbf{M} \times \mathbf{B_0}
\end{align}
where $\gamma$ is the gyromagnetic ratio. $\gamma = 2.675$ \si{\radian.T^{-1}.s^{-1}} for the hydrogen atom ($^1H$) which is the most frequently used atom in MRI as it is omnipresent in the human body. Typically, we are considering an ensemble of spins rather than a single spin with $\mathbf{M}$ then being the sum of magnetisations. In the absence of external forces this will be zero as the spins are pointing in all different directions.

In a static $\mathbf{B_0}$ field, the magnetisation precesses at a frequency dependent on the magnitude of $\mathbf{B_0}$, known as the Larmour frequency: 
\begin{align}
	\omega_0 = \gamma B_0.
\end{align}
With the introduction of a radiofrequency (RF) excitation pulse $\mathbf{B_1}$, $\mathbf{M}$ can be flipped from its equilibrium state $\mathbf{M_0}$ into the transverse plane. The $\mathbf{B_0}$ field is typically thought of as being aligned with the $z$ axis, the longitudinal component, and thus the transverse plane is the $x$\nobreakdash--$y$~plane. The flip angle is given by $\alpha = \gamma B_1 \Delta t$, with $\Delta t$ being the duration of the RF pulse. $\mathbf{M}$ will begin to return to its equilibrium state after the RF pulse is finished and the temporal evolution of $\mathbf{M}$ is given by the Bloch equations:
\begin{align}
	\diff{M_x}{t} & = \omega_0 M_y - \frac{M_x}{T_2} \\
	\diff{M_y}{t} & = - \omega_0 M_x - \frac{M_y}{T_2} \\
	\diff{M_z}{t} & = \frac{M_0 - M_z}{T_1}.
\end{align}
$T_1$ and $T_2$ are the longitudinal and transverse relaxation constants, respectively. The transverse magnetisation decays exponentially with time constant $T_2^*$ and simultaneously, the longitudinal magnetisation returns exponentially to its equilibrium state with time constant $T_1$. $T_2^*$ is a longer decay constant which incorporates the additional transverse decay due to magnetic field inhomogeneities $T_2'$ such that: 
\begin{align}
	\frac{1}{T_2^*} = \frac{1}{T_2} + \frac{1}{T_2'}.
\end{align}	
It is the differences between these relaxation constants, due to the local molecular environment of tissue, that gives different contrasts between tissues and allows us to generate images. The specific sequence of RF pulses and gradient pulses played to achieve the required MR signal is known as the pulse sequence.

\section{Introduction to myocardial perfusion CMR}
The most common approach to the assessment of myocardial perfusion using MRI is with dynamic contrast-enhanced (DCE) acquisitions, as was first described more than 30 years ago \cite{Atkinson1990}. This is done with paramagnetic contrast agents, the most common of which are gadolinium-based. Gadolinium is highly paramagnetic due to its 7 unpaired electrons \cite{Monti2008} and thus the hydrogen nuclei close to the gadolinium will have reduced relaxation times. The $T_1$ relaxation time of the water protons is inversely proportional to the concentration of gadolinium \cite{Salerno2009}: 
\begin{align}\label{eqn:t1_gd}
	R_1 = R_{10} + r_1 [\textnormal{Gd}]
\end{align}
where $R_1 = 1/T_1$ is the longitudinal relaxation rate, $R_{10}$ is the pre-contrast longitudinal relaxation rate, $r_1$ is the relaxivity of the contrast agent ($4.5$ \si{\litre\per\milli\mole\per\second} for gadovist at 3T \cite{Broadbent2016}), and $[\textnormal{Gd}]$ is the concentration of gadolinium.  Thus, areas with a high concentration of gadolinium will appear to be brighter on T1 weighted images. In perfusion CMR, a bolus of contrast agent is administered intravenously and time dynamic images are acquired to observe the temporal evolution of the contrast bolus. Areas of the myocardium with reduced perfusion hence appear hypointense. These perfusion defects correspond to regions of either ischaemia or fibrosis \cite{Plein2015}.

As discussed in Section~\ref{sec:coronaries}, the principle of autoregulation ensures that for even a relatively large coronary artery stenosis there may be no reduction in myocardial blood flow. For this reason, scans are usually acquired at peak vasoldilation such that autoregulation can no longer account for the stenosis. Since exercise is not feasible in the scanner, vasodilator stress is pharmacologically induced. This is typically done with adenosine which causes flow-mediated vasodilation \cite{Lupi1997}. Rest images may also be acquired to check for scar or to account for artefacts. 

\begin{figure}[htbp!]
	\centering
	\includegraphics[trim=15cm 7cm 15cm 5cm,clip=true, angle = 110]{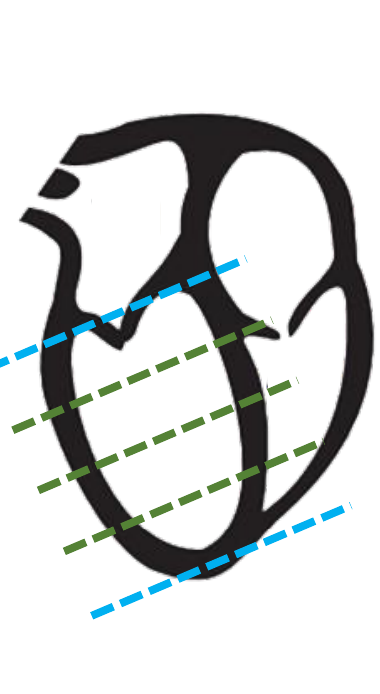}
	\caption["Three of five" slice planning]{Initially five slices with equidistant gaps are planned and the middle three slices (shown in green) are used.}
	\label{fig:3of5}
\end{figure}

To ensure coverage of the LV, 3 slices are typically acquired in a short-axis view using the 'three-of-five' rule \cite{Messroghli2005} where 5 slices are planned with equal slice gap from the base of the LV to the apex and the middle 3 slices are used, demonstrated in Figure~\ref{fig:3of5}. 

The most commonly used pulse sequence for perfusion CMR is a 2D multi-slice saturation recovery (SR) sequence \cite{Plein2015}. A \ang{90} saturation preparation pulse is used to null the pre-contrast signal and to hence visualise the passage of the contrast bolus. Inversion recovery (IR) rather than SR sequences have also been proposed. In this case the \ang{180} preparation pulse inverts the magnetisation to maximise the dynamic contrast-enhancement. The limitations of IR sequences are the increased imaging time makes it incompatible with high heart rates. IR sequences also makes the contrast dependent on the duration of the R-R interval, which is undesirable in patients with varying heart rates, such as in cases of arrhythmia \cite{Kellman2007}. The acquisition is ECG-gated, images are acquired a fixed time (trigger time) after the R wave is recorded to the effect that the slices are always acquired in the same cardiac phase \cite{Plein2015}.

At 3T, the most commonly used readout are spoiled gradient echo readouts. These repeatedly excite the imaging slice with pulses of low flip angle while acquiring k-space data line-by-line. The excitation pulses are with low flip angles to allow rapid imaging, larger flip angles would give more $T_1$ contrast but would not be possible within a single R-R interval. Balanced steady-state free precession (bSSFP) and echo planar imaging (EPI) readouts are also used but are less commmonly employed.

More recently, 3D \cite{Jogiya2012} and simultaneous multi-slice acquisitions \cite{Nazir2018} have been proposed in order to increase the coverage of the LV but these are not widely available and have not yet achieved clinical adoption.

Non-contrast alternatives for myocardial perfusion imaging include 
arterial spin labelled (ASL) and blood oxygen level-dependent (BOLD) CMR. ASL employs RF pulses to locally alter the magnetisation of the arterial blood supply and then images the labelled blood as it reaches the myocardium in order to estimate perfusion \cite{Do2018}. BOLD uses deoxyhaemoglobin as an endogenous contrast agent. Since deoxyhaemoglobin is paramagnetic, it reduces the signal in $T_2$-weighted images and thus gives a direct assessment of myocardial oxygenation and blood flow   \cite{Manka2010}. Both approaches have some limited clinical data supporting their use but they still remain research tools. As such, this thesis will focus solely on DCE perfusion CMR.

\begin{figure}[htbp!]
	\centering
	\includegraphics[width=.9\textwidth]{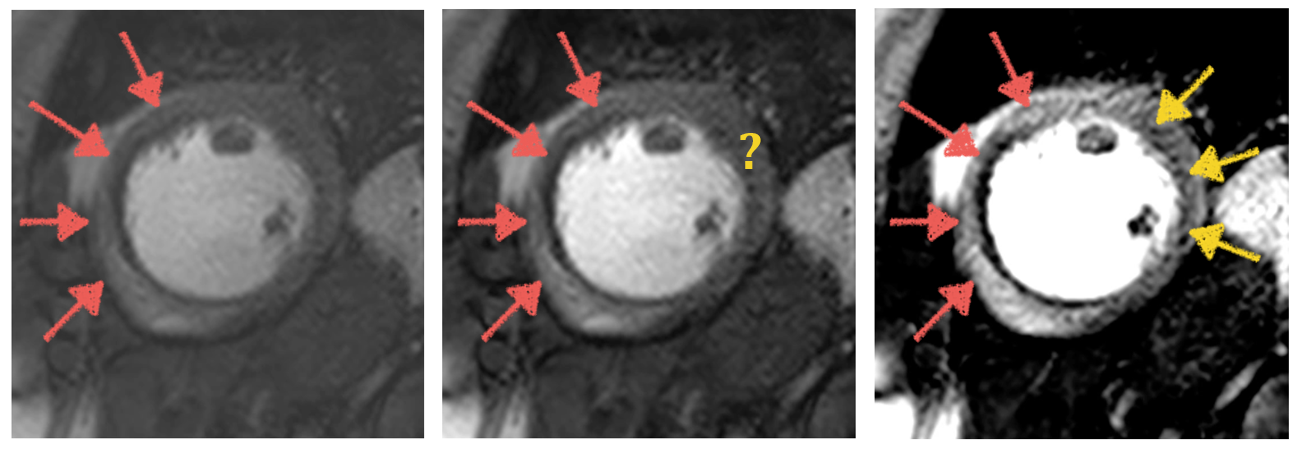}
	\caption[The difficulty of visual assessment]{An example image demonstrating the difficulty of the visual assessment of perfusion CMR. The diagnosis of the patient from one-vessel disease (orange arrows) changes to two-vessel disease (orange and yellow arrows) with the narrowing of the contrast window. At an intermediate level of windowing (centre), the diagnosis is not clear.}
	\label{fig:operators}
\end{figure}

As discussed, due to its high diagnostic accuracy, stress perfusion CMR has become one of the methods of choice for the diagnosis of CAD. It has a class IA recommendation from the European Society of Cardiology (ESC) for the evaluation of patients with an intermediate pretest probability of CAD \cite{members2014}. The main limitation of the modality is the difficulty of interpreting the images. The diagnostic accuracy has been shown to be highly dependent on the level of experience of the operator \cite{Villa2018}. The difficulty of the interpretation of the images is visualised in Figure~\ref{fig:operators} which shows the same image with different levels of contrast windowing. The identification of perfusion defects, as indicated by the arrows, changes with different windowing. A potential solution to this is the quantitative analysis of the images which would add user-independence and reproducibility to the high clinical utility of the modality.

\section{Quantitative myocardial perfusion CMR} \label{sec:qperfCMR}
The goal of quantitative myocardial perfusion CMR is to infer the kinetics of the myocardium from the observed signal (contrast) evolution. The quantification of these tissue properties give a high level of diagnostic information about the patient. 

\subsection{Signal intensity to contrast concentration conversion}

The first challenge in this process is the conversion of the MR signal intensities (SI) to the concentration of the contrast agent, gadolinium [Gd] (in units of \si{\Molar}). In the ideal case, there is a linear relationship between SI and [Gd]. As will be shown in Section~\ref{section:tkmodel}, the tracer-kinetics are modelled as a linear time-invariant system so that, in the case of linear relationship, the parameter estimates derived with the SI curves directly are equal to those derived with the concentration curves. The reality, however, is that there is a more complex, non-linear relationship between SI and [Gd] \cite{Bokacheva2007} causing a signal saturation for high [Gd] \cite{Ishida2011}. As a result, the relationship between SI and [Gd] needs to be modelled. This is done by estimating $T_1$ at each time from the corresponding SI and relating $T_1$ to [Gd] through equation~\ref{eqn:t1_gd}. The signal equation, as a function of $T_1$, for a saturation-recovery spoiled gradient echo sequence, as commonly used in perfusion CMR, is given as:
\begin{align} \label{eqn:si}
	S = \psi S_0 \bigg[ (1 - \exp(-T_{SAT} / T_1 )) a^{n-1} + (1 - \exp(-T_R / T_1 )) \frac{1-a^{n-1}}{1-a}  \bigg]
\end{align}
where $\psi$ is a constant scaling factor that absorbs factors such as the coil sensitivities and system gains, $S_0$ is the baseline signal level, $T_{SAT}$ is the time between the saturation preparation pulse and the acquisition of the central line of k-space, $T_R$ is the repetition time between excitation pulse, $a = \cos\alpha \exp(-T_R / T_1)$, $\alpha$ is the flip angle and $n$ is the number of excitation pulse from the beginning until the centre of k-space \cite{Bokacheva2007}. $\psi$ is assumed to be constant over time and can be estimated from the baseline pre-contrast images using a baseline $T_{10}$. Since $S = \psi f(T_1)$, the estimated $\psi$ is given as $\psi = f(T_{10}) / S_0$. The baseline $T_{10}$ can either be taken from literature values or from a pre-contrast $T_1$ map. The $T_1$ value at each time point is then determined using a root-finding algorithm.

\subsection{Tracer-kinetic modelling}
\label{section:tkmodel}
In theory, the gadolinium-based contrast agents used in perfusion CMR are indicators rather than tracers as they are not chemically the same as the systemic substance of interest \cite{Sourbron2011}. Nonetheless, the word tracer will be used in order to maintain consistency with the literature.

\subsubsection*{General theory}
A tissue is modelled as system with a series of inlets and outlets through which the system substance can flow. The models are built on the theory of linear time-invariant systems \cite{Sourbron2013}. That is, that the transit time, the time elapsed between entering and leaving the system, does not depend on the time of the contrast injection or the injected concentration. The system is governed by the conservation of mass. This states that no tracer is created or destroyed in the system and gives that the rate of change of concentration in the tissue is the difference between the influx and outflux through the inlets and outlets of the system \cite{Sourbron2013}:
\begin{align}\label{eqn:orig_ode}
	v \frac{dC(t)}{dt} = \sum_{i} F_i C_i(t) - \sum_{o} F_o C_o(t)	
\end{align}
where $v$ is volume of distribution (the volume of the system that contains the tracer), $C(t)$ is the total concentration of tracer in the system and, $F_i, F_o$ and $C_i(t), C_o(t)$ are the flows through and concentrations at the inlets $i$ and outlets $o$ of the system. The system itself can be made up of a number of interacting compartments with the inlets of a compartment possibly being the outlets of another compartment. Equation~\ref{eqn:orig_ode} can be applied to each compartment to yield a system of $n$ ordinary differential equations (ODEs).

\subsubsection*{Two-compartment exchange model}
In the myocardium, there is a single arterial input and single outlet. The tracer is assumed to be contained in the blood plasma and the extravascular extracellular space (EES) and that each of these spaces are an individual compartment. This gives rise to, under some further assumptions, to the two-compartment exchange model (2CXM) \cite{Jerosch-herold2010}. The further assumptions made are that the influx and outflux of the system is through the plasma compartment, the EES compartment only exchanges with the plasma compartment, and that this exchange is equal in both directions. \begin{figure}[htbp!]
	\centering
	\includegraphics[width=.9\textwidth]{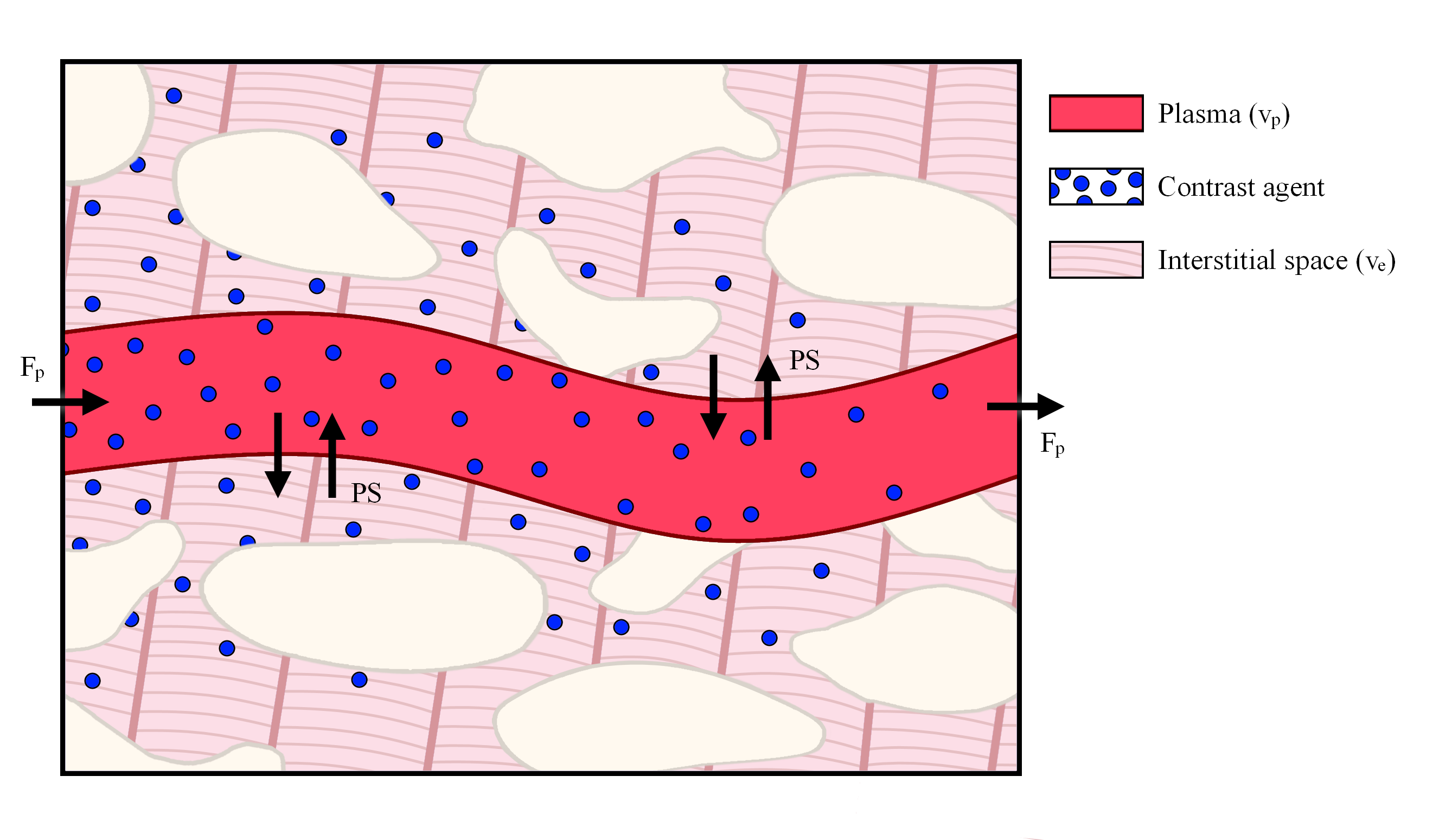}
	\caption[Two-compartment exchange model]{A visualisation of the underlying physiological representation of the 2CXM.}
	\label{fig:2cxm}
\end{figure} This then gives two coupled ODEs for the concentration of contrast in the plasma $C_p$ (\si{\Molar}) and EES $C_e$ (\si{\Molar}) compartments \cite{Sourbron2013}:
\begin{align}
	v_p \frac{dC_p(t)}{dt} & = F_p \cdot (C_{AIF}(t) - C_p(t)) + PS \cdot (C_e(t) - C_p(t)) \label{eqn:2cxm1} \\
	v_e \frac{dC_e(t)}{dt} & = PS \cdot (C_p(t) - C_e(t)). \label{eqn:2cxm2}
\end{align}
In these equations, $C_{AIF}(t)$ (\si{\Molar}) is the arterial input function (AIF), the assumed input to the system that is being modelled. $F_p$ is the plasma flow (\si{\milli\litre\per\min\per\milli\litre}) , $v_p$ is the fractional plasma volume (dimensionless), $v_e$  is the fractional interstitial volume (dimensionless) and PS is the permeability-surface area product (\si{\milli\litre\per\min\per\milli\litre}). The total concentration of contrast in the myocardial tissue is then given as the weighted sum of the concentration in the compartments:
\begin{align}
C_{myo}(t) = v_pC_p(t) + v_eC_e(t).
\end{align}

The 2CXM can be modified by making additional assumptions that the system contains only one compartment or that there is uptake of contrast rather than exchange of contrast in the tissue \cite{Sourbron2013}. It can further be extended to include a spatial component, i.e to assume that the concentration of contrast is not uniformly distributed over a compartment. 

In this thesis, the 2CXM will be the focus as it balances most closely matching the underlying physiology while still having an analytic solution.

\subsubsection*{Solution of the two-compartment exchange model}
\begin{figure}[htbp!]
	\centering
	\includegraphics[trim=2cm 9cm 2cm 5cm,clip=true,width=\textwidth]{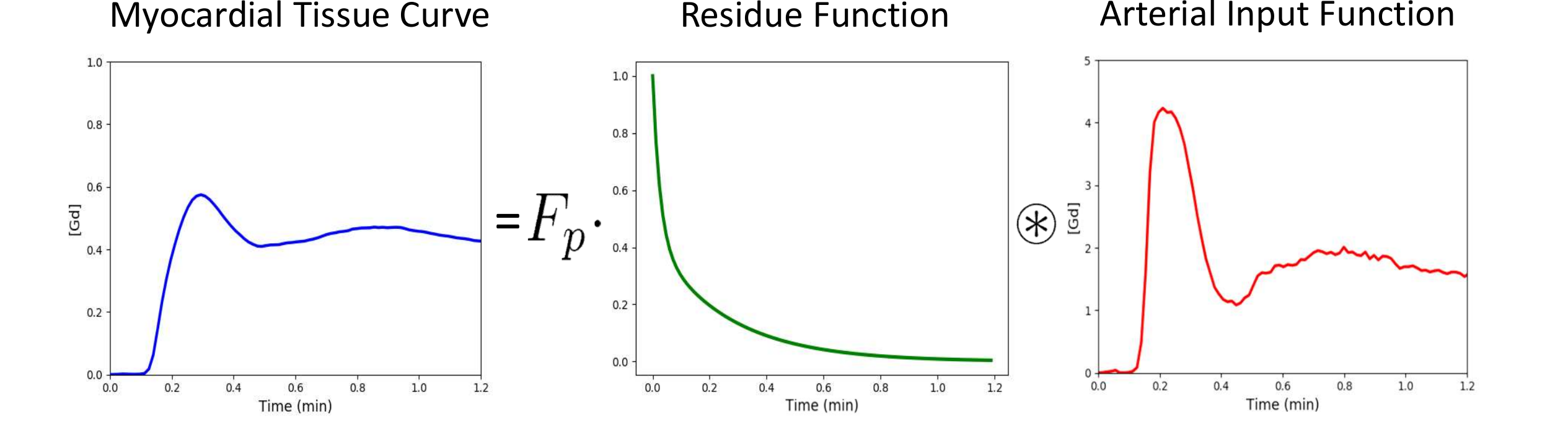}
	\caption[The relation of myocardial tissue curve to AIF]{The myocardial tissue curves is shown as the result of a convolution of the AIF with a residue curve.}
	\label{fig:residue_conv}
\end{figure}

An analytic solution for $C_{myo}(t)$ can be obtained using the Laplace transform under the assumption of zero initial concentration $C_p(t=0)=0$ and $C_e(t=0)=0$. In the Laplace domain, the coupled ODEs in equations~\ref{eqn:2cxm1} and ~\ref{eqn:2cxm2}  become:
\begin{align}
	v_p \tilde{C_p}(s) & = F_p \cdot (\tilde{C}_{AIF}(s) - \tilde{C_p}(s)) + PS \cdot (\tilde{C_e}(s) - \tilde{C_p}(s)) \label{eqn:l2cxm1} \\
	v_e \tilde{C_e}(s) & = PS \cdot (\tilde{C_p}(s) - \tilde{C_e}(s)) \label{eqn:l2cxm2}
\end{align}
where $\tilde{C}(s)$ is the Laplace transform of $C(t)$. Isolating $\tilde{C_e}$ from equation~\ref{eqn:l2cxm2} yields:
\begin{align}
	\tilde{C_e}(s) = \frac{PS}{v_e s + PS}	\tilde{C_p}(s)
\end{align}
which can be substituted back into equation~\ref{eqn:l2cxm1} to give:
\begin{align}
	(v_p s + F_p + PS) \tilde{C_p}(s) = F_p \cdot \tilde{C}_{AIF}(s) + PS \frac{PS}{v_e s + PS} \tilde{C_p}(s)
\end{align}
and finally: 
\begin{align}
	\tilde{C_p}(s) = F_p \cdot \tilde{C}_{AIF}(s) \frac{\tfrac1{v_p} (s + \tfrac{PS}{v_e}) }{s^2 + s (\tfrac{F_p}{v_p}+ \tfrac{PS}{v_p}+\tfrac{PS}{v_e}) + \tfrac{F_p}{v_p} \tfrac{PS}{v_e}}.
\end{align}
Similarly, 
\begin{align}
	\tilde{C_e}(s) = F_p \cdot \tilde{C}_{AIF}(s) \frac{\frac{PS}{v_p v_e}}{s^2 + s (\tfrac{F_p}{v_p}+ \tfrac{PS}{v_p}+\tfrac{PS}{v_e}) + \tfrac{F_p}{v_p} \tfrac{PS}{v_e}}.
\end{align}
Hence, the total tissue concentration in the myocardium can be written in Laplace space as:
\begin{align} 
	\tilde{C}_{myo}(s) & = v_p \tilde{C_p}(s) + v_e \tilde{C_e}(s) \\
	& = F_p \cdot \tilde{C}_{AIF}(s) \frac{s + \tfrac{PS}{v_p} + \tfrac{PS}{v_e}}{s^2 + s (\tfrac{F_p}{v_p}+ \tfrac{PS}{v_p}+\tfrac{PS}{v_e}) + \tfrac{F_p}{v_p} \tfrac{PS}{v_e}}. \label{eqn:lap_soln}
\end{align}
It is observed that the denominator of equation~\ref{eqn:lap_soln} is a quadratic function of $s$ and that its roots are given by:
\begin{align} 
	\begin{pmatrix}
		\alpha \\
		\beta
	\end{pmatrix}	
		 =  \bigg[ - \tfrac12 \bigg(\tfrac{F_p}{v_p}+ \tfrac{PS}{v_p}+\tfrac{PS}{v_e}\bigg) \pm \sqrt{\bigg(\tfrac{F_p}{v_p}+ \tfrac{PS}{v_p}+\tfrac{PS}{v_e}\bigg)^2 - 4 \tfrac{PS}{v_e}\tfrac{F_p}{v_p}} \ \bigg]
\end{align}
and so by the use of partial fractions equation~\ref{eqn:lap_soln} can be rewritten as: 
\begin{align} 
	\tilde{C}_{myo}(s) = F_p \cdot \tilde{C}_{AIF}(s) \bigg( \frac{A}{s-\alpha} + \frac{1-A}{s-\beta} \bigg)
\end{align}
with $A = \frac{\alpha + \tfrac{PS}{v_p}+\tfrac{PS}{v_e} }{\alpha - \beta}$. The inverse Laplace transform then yields the time domain solution:
\begin{align} \label{eqn:conv_soln}
	C_{myo}(t) = F_p \cdot R(t) \ast C_{AIF}(t)
\end{align}
with:  
\begin{align} 
	R(t) = A \exp( \alpha t) + (1 - A) \exp(\beta t)
\end{align}
being the residue function which can also be written as $R(t;\boldsymbol{\theta})$ to emphasise its dependence on the kinetic parameters $\boldsymbol{\theta} = (F_p, v_p, v_e, PS)$. The solution given by Equation~\ref{eqn:conv_soln} is shown in terms of the concentration curves in Figure~\ref{fig:residue_conv}.

The model, as derived, considers the plasma compartment and hence the plasma flow and plasma volume parameters. Since the convention is to report blood flow and blood volume parameters a conversion can be made using the blood haematocrit value $F_b = F_p / (1-\textnormal{Hct})$ and $v_b = v_p / (1-\textnormal{Hct})$. The AIF, which is sampled from the LV cavity, is converted from arterial blood concentration to arterial plasma concentration by substituting $C_{AIF}(t) / (1-\textnormal{Hct})$ for $C_{AIF}(t)$. Patient specific haematocrit values could be used but they are not typically available and a literature reference of Hct $= 0.45$ is used \cite{Broadbent2016}. Flow values are converted from units of \si{\milli\litre\per\min\per\milli\litre} to its mass equivalent \si{\milli\litre\per\min\per\gram} by multiplying by the specific density of the myocardium, taken as $1.05$ \si{\gram\per\milli\litre}. Note that $F_b$ is often referred to as myocardial blood flow and will be used interchangeably with the acronym MBF.

\subsubsection*{Kinetic parameter estimation} \label{sec:kin_param}
The most common approach to the estimation of the model parameters from observed myocardial and arterial concentration curves is using non-linear least squares fitting \cite{Ahearn2005} with the most popular choice of fitting algorithm being the Levenberg–Marquardt approach \cite{Marquardt1963}. That is to say, the sum of squared errors between the analytic solution for $C_{myo}$ and the observed myocardial tissue concentration curves is minimised with respect to $\boldsymbol{\theta}$. The parameters $\boldsymbol{\hat{\theta}}$ giving the optimal sum of squared errors are taken as an estimate of the true parameters, this is:
\begin{align}
	\boldsymbol{\hat{\theta}} = \argmin_{\boldsymbol{\theta}} \| C_{myo}(t) - F_p \cdot R(t;\boldsymbol{\theta}) \ast C_{AIF}(t)  \|_2^2.
\end{align}
This process is sometimes (incorrectly) referred to as a deconvolution.

The ability to accurately estimate $\boldsymbol{\theta}$ can depend on the quality of the data. The data quality can be limited by the signal-to-noise ratio (SNR), the acquisition time, the temporal resolution, and artefacts, such as motion \cite{Ahearn2005}. These introduce local optima in the cost function. This means that there are distinct sets of parameters that are indistinguishable at the noise level present in the data \cite{Buckley2002}. The result of this is that the parameter estimates obtained from the non-linear least squares fitting algorithm are highly dependent on the initial starting point of the optimisation. Ahearn et al. \cite{Ahearn2005} reported that repeating the optimisation with multiple different starting points can improve the reliability but still is far from perfect even using simulated data without noise. Buckley \cite{Buckley2002} found an array of local minima in which different combinations of parameters give very similar solutions, while Jerosch-Herold et al. \cite{Jerosch-Herold2002} were unable to get unique estimate for $F_p$.

The reliability of the estimated parameters can also be improved by reducing the complexity of the model used \cite{Buckley2002}. As mentioned, other choices of model are possible. Examples of models commonly employed for myocardial perfusion quantification are the Kety-Tofts model \cite{Likhite2017} and the Fermi model \cite{Jerosch-Herold2002}. These models have less parameters to estimate, making the fitting problem more stable. However, they have limitations in that the Kety-Tofts model does not resolve directly for $F_p$ and the Fermi model is not physiologically motivated. Alternatively, approaches for improving the parameter estimates with the 2CXM include fitting to a concentration curve averaged over a whole segment of myocardium. This sacrifices resolution in order to increase contrast-to-noise (CNR). However, it is well known that high resolution maps are needed for detecting subtle ischaemia \cite{Le2020}.

This motivates the need for more robust fitting approaches, as will be discussed in Chapter 6.

\subsubsection*{Arterial input function}
\begin{figure}[htbp!]
	\centering
	\includegraphics[width=.9\textwidth]{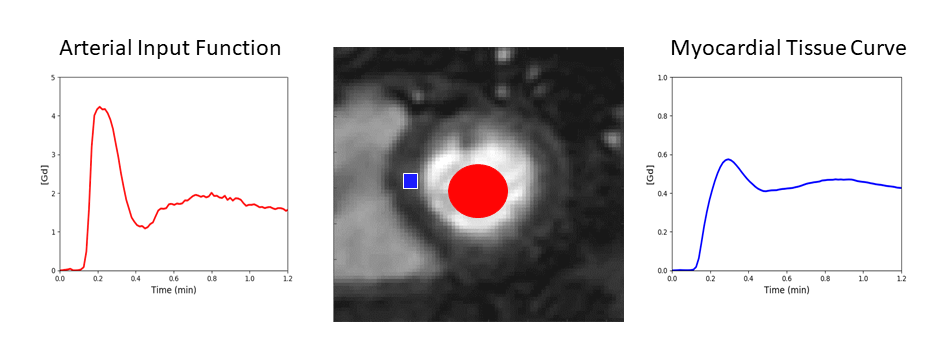}
	\caption[The AIF and myocardial tissue curve]{The AIF (red, left) and myocardial tissue curve (blue, right) shown with their corresponding sampling locations overlaid on the MR image (centre).}
	\label{fig:aif_curve}
\end{figure}

The accurate estimation of the AIF is one of the key challenges of myocardial perfusion quantification. In theory, the AIF should be sampled from the inlet of the system. In the case of myocardial perfusion CMR this would be the root of the aorta. However, this is not typically imaged in a standard perfusion CMR acquisition and those the AIF is sampled from the LV. An example image with the sampling locations of is shown in Figure~\ref{fig:aif_curve}. The effect of sampling in the aorta rather than the LV would be a delay and dispersion in the AIF \cite{Calamante2000}. The delay can be accounted for by replacing $C_{AIF}(t)$ with $C_{AIF}(t-\tau)$ where $\tau$ is the delay and can either be fit as an extra model parameter or estimated separately. The dispersion could also be modelled and accounted for \cite{Calamante2000}, though this is not done in perfusion CMR and leads to a systematic under-estimation of perfusion.

Since the whole bolus of contrast passes through the LV cavity more-or-less simultaneously, very high concentrations of contrast agent are recorded at the peak of the AIF. As a result, the relationship between the signal intensity and contrast concentration in the AIF is non-linear \cite{Sanchez-Gonzalez2015}. Conversely, as only a small fraction of the total amount of the contrast inject perfuses a particular area of myocardium, the concentration remains low and a linear relationship between signal intensity and contrast concentration is assumed to hold.

As discussed, the contrast concentration can be estimated from the AIF SI using Equation~\ref{eqn:si}. However, this fitting suffers from the same difficulties that were discussed in Section~\ref{sec:kin_param} and it would beneficial if a linear relationship could be assumed so that the tracer-kinetic modelling could be performed directly using the relative signal enhancement approximation to the contrast concentration \cite{Biglands2015}:
\begin{align}
	C(t) \approx \frac{R_{10}}{r_1} \cdot \frac{S(t) - S(0)}{S(0)}.
\end{align}

For this reason, a significant amount of effort and research in methods for myocardial perfusion quantification is focused on approaches for approximating a linear relationship between signal intensity and contrast concentration in the LV cavity. There are currently two approaches for doing this: the dual-bolus approach and the dual-sequence approach. 

The dual-bolus approach uses a low-dose bolus (the pre-bolus) of contrast agent (typically $1/10$\textsuperscript{th} of the full dose) in attempt to avoid the signal saturation at high contrast concentrations \cite{Ishida2011}. This low dose, for AIF estimation, does not yield enough signal to assess the myocardium so it is then followed by a full dose (the main bolus). For the tracer-kinetic modelling, the AIF from the pre-bolus is scaled up (by 10) to approximate the main bolus without signal saturation and used with the myocardial curves from the main bolus. The dual-bolus method is well validated and has proven to give accurate estimates of perfusion \cite{Hsu2012,Morton2012a,Schuster2015}. However, its clinical adoption has been limited due to the complexity added to the scan and the extra work involved in the two injections.

The dual-sequence method acquires an extra low resolution image slice with a short saturation-recovery time to minimise the saturation of the AIF signal \cite{Gatehouse2004}. The benefit of the dual-sequence is that perfusion can be quantified accurately with a single bolus of contrast. The clinical adoption has been slow as the sequence has yet to be commercialised and made available on all scanners. However, recent implementations of research prototypes are making the sequence more widely available \cite{Sanchez-Gonzalez2015, Kellman2017}. The limitation is that the extra image slice adds additional time to the acquisition, at high heart-rates it, therefore, may not be possible to acquire all slices in one cardiac phase and the spatial resolution may have to be compromised. Furthermore, while the short saturation time minimises the signal saturation, it may not completely prevent it.

Recent work has demonstrated that since there is an extra image required for the dual-sequence approach that it is feasible to not collocate this with the myocardial slices \cite{Mendes2020}. Mendes et al. chose to place the AIF slice in the ascending aorta which should be more accurate in theory, though further research is required to fully explore this direction.

\section{Literature review}

There is a wealth of literature on methods, evaluation, and validation of quantitative perfusion CMR. A lot of the early work focused on the use of semi-quantitative metrics where, in lieu of the full tracer-kinetic modelling, the ratio of the up-slope in the myocardial tissue curves to the up-slope in the AIF is used as a surrogate for flow. This has been validated versus coronary angiography \cite{Nagel2003,Giang2004,Plein2005} and FFR \cite{Rieber2006} with favourable accuracy as compare to visual assessment. It has been further compared to the non-invasive reference standard for perfusion quantification, PET, showing a good correlation between the methods \cite{Schwitter2001, Ibrahim2002}. Al-Saadi et al. \cite{Al-Saadi2000} used a training set to derive a cutoff for ischaemia and applied the threshold obtained prospectively to a new cohort and found a diagnostic accuracy of $87\%$ for the detection of a significant stenosis on the coronary angiogram.

The limitation of the semi-quantitative measures is that they are not physiological and as such are subject to variations across patients, scanners, and implementations. The wide variations in absolute values motivated the use of the myocardial perfusion reserve (MPR) instead of the absolute values. The MPR is the ratio of flow at stress to flow at rest and the idea is that the variations may cancel themselves out in the ratio. This is not really the case and, indeed, Mordini et al. \cite{Mordini2014} showed superior diagnostic performance for fully-quantitative perfusion over the semi-quantitative measures. Furthermore, there is a recent trend towards stress only imaging to reduce scan time and thus MPR is no longer possible and absolute quantification of stress perfusion is desired.

Other semi-quantitative measure have also been developed. Hautvast et al. \cite{Hautvast2011} computed gradients in the signal-intensity curves in the transmural direction, across the myocardial wall. This built on the knowledge that perfusion defects appear earlier and more severely in the sub-endocardial wall. Chiribiri et al. \cite{Chiribiri2016}. used the temporal dyssynchrony of flow in the myocardium. This is effective at detecting CAD as a coronary artery stenosis causes temporal delays and dyssynchrony in the contrast flow, while contrast flow is temporally uniform in the absence of a stenosis. While both approaches showed promising performance, neither has seen widespread adoption.

More recently, much of the focus in the field has been towards fully-quantitative perfusion CMR. This has again been compared extensively to PET perfusion estimates \cite{Fritz-Hansen2008a, Morton2012, Miller2014}. All of these studies found a strong linear correlation even if the absolute values do not agree perfectly due to the differences in the quantification methods. Similarly, two studies have found a high correlation between perfusion quantified by CMR and by fluorescent microspheres in dogs \cite{Hsu2012} and pigs \cite{Schuster2015}. This indicates that quantitative perfusion CMR is indeed accurately estimating flow and paves the way for its use in the clinic. 

There has been further validation versus invasive measurements in patients with CAD. Lockie et al. \cite{Lockie2010} reported an area under the curve (AUC) of 0.89 on the receiver operating characteristic (ROC) analysis versus FFR. This matched the accuracy of the expert visual assessment. Similar accuracy in comparision to invasive measurements was reported in a series of further studies \cite{Biglands2015,Papanastasiou2016,Biglands2018,Hsu2018}. Out of these, importantly, Papanastasiou et al. \cite{Papanastasiou2016} reported superior diagnostic accuracy using a full tracer-kinetic model rather than the Fermi function approximation and they also, along with Biglands et al. \cite{Biglands2018}, found no benefit of including the rest results in addition to the stress perfusion values. Sammut et al. \cite{Sammut2017} took all of this a step further and showed the prognostic value and in particular that the ischaemic burden computed by quantitative perfusion CMR predicts adverse cardiac events.

The focus on quantitative perfusion may seem disproportionate considering than none of the aforementioned studies show that it outperforms the visual assessment. However, it should be noted that all of the studies were conducted at very experienced research hospitals. Villa et al. \cite{Villa2018} showed that the performance of the visual assessment drops off quickly for less experienced operators. The hypothesis is that quantitative perfusion is less user-dependent than the visual assessment that it could more easily generalise to less experienced centres.

The drawback of all the studies is that the quantification still involved tedious manual interaction for the selection of the AIF, segmentation of the myocardium, motion correction, and time delay estimation. For example, Biglands et al. \cite{Biglands2018} reported that it took one hour per patient to manually correct the myocardial segmentation to fit the motion in the data which is not feasible in clinical routine. This motivates the aim of this thesis, to build and validate a fully-automatic pipeline including robust motion compensation and more reliable estimates of the kinetic parameters, negating some of the problems discussed in Section~\ref{section:tkmodel}.

There has been other work developed concurrently or after the work presented in this thesis. Xue et al. \cite{Xue2020} presented an automated pipeline using their dual sequence implementation \cite{Kellman2017}. This was compared to PET in a semi-automated approach which automatically generated pixel-wise flow maps but required manual segmentation of myocardium and its sub-segmentation \cite{Engblom2017}. Finally, they also demonstrated the prognostic value of quantitative perfusion CMR in a study which included automatic segmentation \cite{Knott2020}.

\chapter[Motion compensation]{Robust non-rigid motion compensation of free-breathing myocardial perfusion MRI data}

\section{Preface}

\begin{figure}
	\centering
	\includegraphics[trim=2cm 6cm 2cm 6.5cm,clip=true,width=\textwidth]{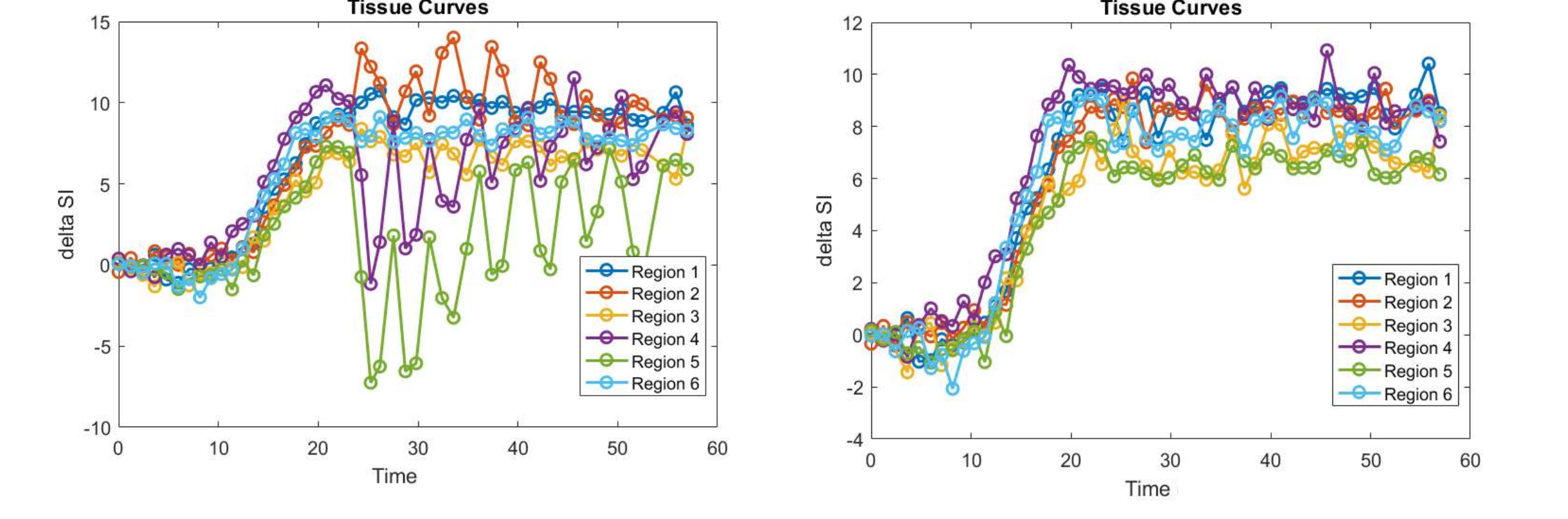}
	\caption[The motion profile of myocardial tissue curves before and after motion compensation]{The motion profile of myocardial tissue curves before and after motion compensation taken as the average signal from each of the 6 AHA segments of one myocardial slice.}
	\label{fig:motion_curves}
\end{figure}

The reliable quantification of myocardial perfusion on a pixel-wise level assumes that the myocardium remains stationary over time. This is of course not realistic as the heart is contracting and the patient is breathing. Cardiac motion is well accounted for in perfusion CMR as the acquisition are ECG-triggered so that a slice is always imaged in the same cardiac phase. However, respiratory motion can cause inter-frame misalignment. The effect of this inter-frame misalignment on the myocardial tissue curves is shown in Fig~\ref{fig:motion_curves} which can lead to errors in the tracer-kinetic modelling.

The most common way to minimise respiratory motion is breath-holding but since typical scans are longer than a minute, the breath-hold cannot cover the full scan. This is especially true for patients with heart disease who have trouble breathing and breath-holding. Long breath-holds also induce changes in heart rate and, thus, cause images to be acquired at different cardiac phases \cite{Pontre2017}. For visual assessment, a short breath-hold around the time of the first-pass of the contrast bolus through the myocardium is sufficient. However, this is not the case for tracer-kinetic model fitting, particular for accurately identifying the microvascular kinetics which play out at longer time scales.

Motion compensation techniques are based on using image registration to correct the inter-frame misalignment. Mathematically, this is to define a cost function $C$ describing how dissimilar an image $I_i$ is from a reference image $R_i$ and to find the transformation $T_i$ from the set of possible transformations $\Omega$ that minimises this cost:
\begin{align}
	\hat{T}_i = \argmin_{T_i \in \Omega} \ \big[C\big(T_i(I_i), R_i \big) + \kappa E(T_i)\big] \ \ \forall \ i=1,2,\ldots,N 
\end{align}
where $N$ is the total number of images and $E$ is a regularisation term, to enforce smooth transformations, controlled by the parameter $\kappa$. 

\begin{figure}
	\centering
	\includegraphics[width=.9\textwidth]{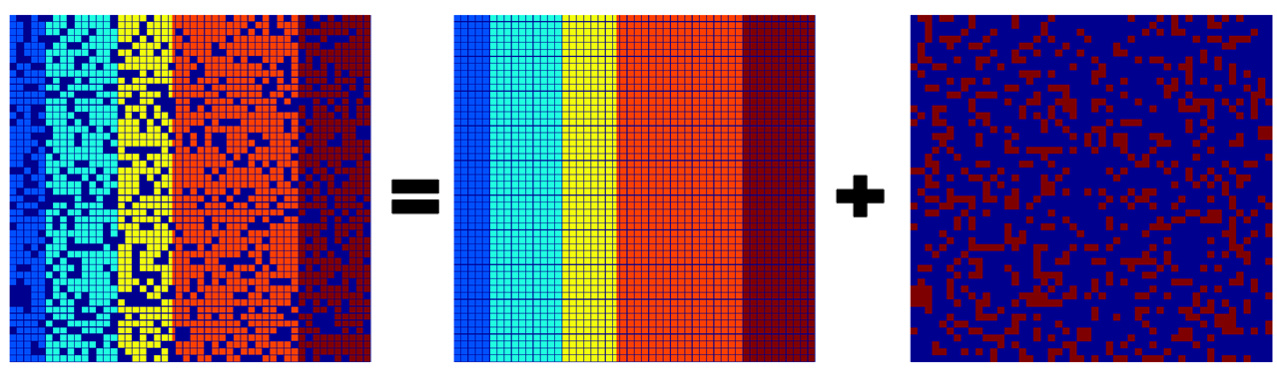}
	\caption[An example of a RPCA decomposition applied to a toy problem]{An example of a RPCA decomposition applied to a toy problem, where it is seen to decompose a noise corrupted matrix into its constituent low-rank (structure) and sparse (noise) components \cite{Perception2017}.}
	\label{fig:rpca_eqn}
\end{figure}

\begin{figure}
	\centering
	\includegraphics[width=.9\textwidth]{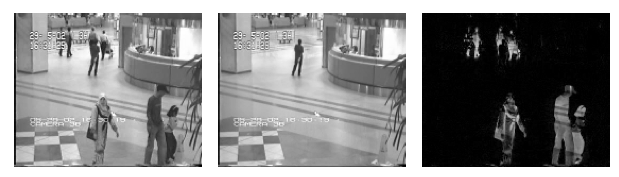}
	\caption[RPCA applied to video surveillance data.]{RPCA applied to video surveillance data, where it is seen that the low-rank component well models the background and the sparse component models the foreground \cite{Perception2017}.}
	\label{fig:video_sur}
\end{figure}

The difficulty of applying image registration to perfusion CMR images is the dynamic contrast-enhancement. The typical cost functions cannot disentangle whether the dissimilarity in two images is cause by inter-frame misalignment or just by the contrast agent being in a different location. This work uses robust principal component analysis, which decomposes a corrupt matrix $\mathbf{M}$ into its constituent low-rank $\mathbf{L}$ and sparse $\mathbf{S}$ components such that $\mathbf{M} = \mathbf{L} + \mathbf{S}$, to separate the dynamic contrast-enhancement from the baseline signal. An example RPCA decomposition for a toy example from \cite{Perception2017} is shown in Figure~\ref{fig:rpca_eqn}. A more interesting example is seen in Figure~\ref{fig:video_sur} (from \cite{Perception2017}) where RPCA is applied to video surveillance data, it is seen that it well decomposes the video into background (this is low-rank as it remains constant over time) and foreground (this is sparse as it is constantly changing).
 
It is perhaps somewhat surprising but such a separation can be easily computed \cite{Candes2009} and it can be formulated as the solution to:  
\begin{align}
	\hat{\mathbf{L}},\hat{\mathbf{S}} = \argmin_{\mathbf{L},\mathbf{S}} \ \  \textnormal{rank}(\mathbf{L}) + \lambda \|\mathbf{S}\|_0 \ \ \textnormal{s.t} \ \ \mathbf{L} + \mathbf{S} = \mathbf{M}.
\end{align}
\vspace{-2mm}
Since the $l_0$ norm, which counts nonzero entries, cannot be easily minimised, it is replaced by the $l_1$ norm, which also encourages sparseness \cite{Natarajan1995}, leading to:
\begin{align}
	\hat{\mathbf{L}},\hat{\mathbf{S}} =     \argmin_{\mathbf{L},\mathbf{S}} \ \  \|\mathbf{L}\|_* + \lambda \|\mathbf{S}\|_1 \ \ \textnormal{s.t} \ \ \mathbf{L} + \mathbf{S} = \mathbf{M}.
\end{align}	
where $\|\cdot\|_*$ is the sum of the singular values of the matrix and is known as the nuclear norm. The optimisation is solved using augmented Lagrangian multipliers, in an alternating directions manner \cite{Lin2011}. That is that we iteratively solve the problem for $\mathbf{L}$ and $\mathbf{S}$ rather than for both simultaneously. This, intuitively, alternates between soft-thresholding the singular values of $\mathbf{L}$ to encourage low-rankness and soft-thresholding the entries of $\mathbf{S}$ to encourage sparseness. With the soft-thresholding operator defined as: 
\begin{align}
	\big(\mathcal{S}_\epsilon [\mathbf{X}] \big)_{ij} := 
	\begin{cases} 
		X_{ij} - \epsilon & \textnormal{if} \ \ X_{ij}\geq \epsilon \\
		X_{ij} + \epsilon & \textnormal{if} \ \ X_{ij} < \epsilon
	\end{cases}
\end{align}
this yields the alternating direction method of multipliers (ADMM) for RPCA in Algorithm~\ref{algo:admm}.\\

\begin{algorithm}[H] \label{algo:admm}
	\SetKwInOut{Input}{Input}
	\SetKwInOut{Output}{Output}
	\SetKwInOut{Initialise}{Initialise}
	
	\Input{ $\mathbf{M}$, $\lambda > 0$}
	\Initialise{ $k=0$, $\mathbf{S}_0=\mathbf{Y}_0=0$, $\mu>0$}
	
	\While{not converged}{
		$(\mathbf{U},\boldsymbol{\Sigma},\mathbf{V}) \leftarrow \textnormal{svd}(\mathbf{M}-\mathbf{S}_k+\mu^{-1} \mathbf{Y}_k)$ \;
		$\mathbf{L}_{k+1} \leftarrow \mathbf{U} \mathcal{S}_{\mu^{-1}}[\boldsymbol{\Sigma}] \mathbf{V}^T$ \;
		$\mathbf{S}_{k+1} \leftarrow \mathcal{S}_{\lambda \cdot \mu^{-1}}[\mathbf{M}-\mathbf{L}_{k+1}+\mu^{-1}\mathbf{Y}_k]$ \;
		$\mathbf{Y}_{k+1} \leftarrow \mathbf{Y}_k + \mu (\mathbf{M} - \mathbf{L}_{k+1} - \mathbf{S}_{k+1})$ \;
	}
 	\Output{ $\hat{\mathbf{L}} = \mathbf{L}_{k}$, $\hat{\mathbf{S}} = \mathbf{S}_{k}$}
	\caption{RPCA via ADMM}
\end{algorithm}
\vspace{3mm}
\noindent where svd is the singular value decomposition.

There has been a wide variety of approaches published in the literature in attempt to solve this problem, as described in part II B of Section~\ref{section:moco_paper} and in the recent benchmark paper by Pontre et al. \cite{Pontre2017}. However, none have gained widespread adoption or are used in clinical practice, possibly due to a lack of validation and robustness, motivating the need for our reliable motion compensation scheme.

\section{Journal article} \label{section:moco_paper}
The following text is reproduced as published \cite{Scannell2019b}:

\vspace{3mm}

\noindent \textit{\textbf{Scannell, C.M.}, Villa, A.D.M., Lee, J., Breeuwer, M. \& Chiribiri, A. Robust Non-Rigid Motion Compensation of Free-Breathing Myocardial Perfusion MRI Data. IEEE Trans. Med. Imaging 38, 1812–1820 (2019).}
:
\newpage
\includepdf[pages=-,pagecommand={},width=1.15\textwidth]{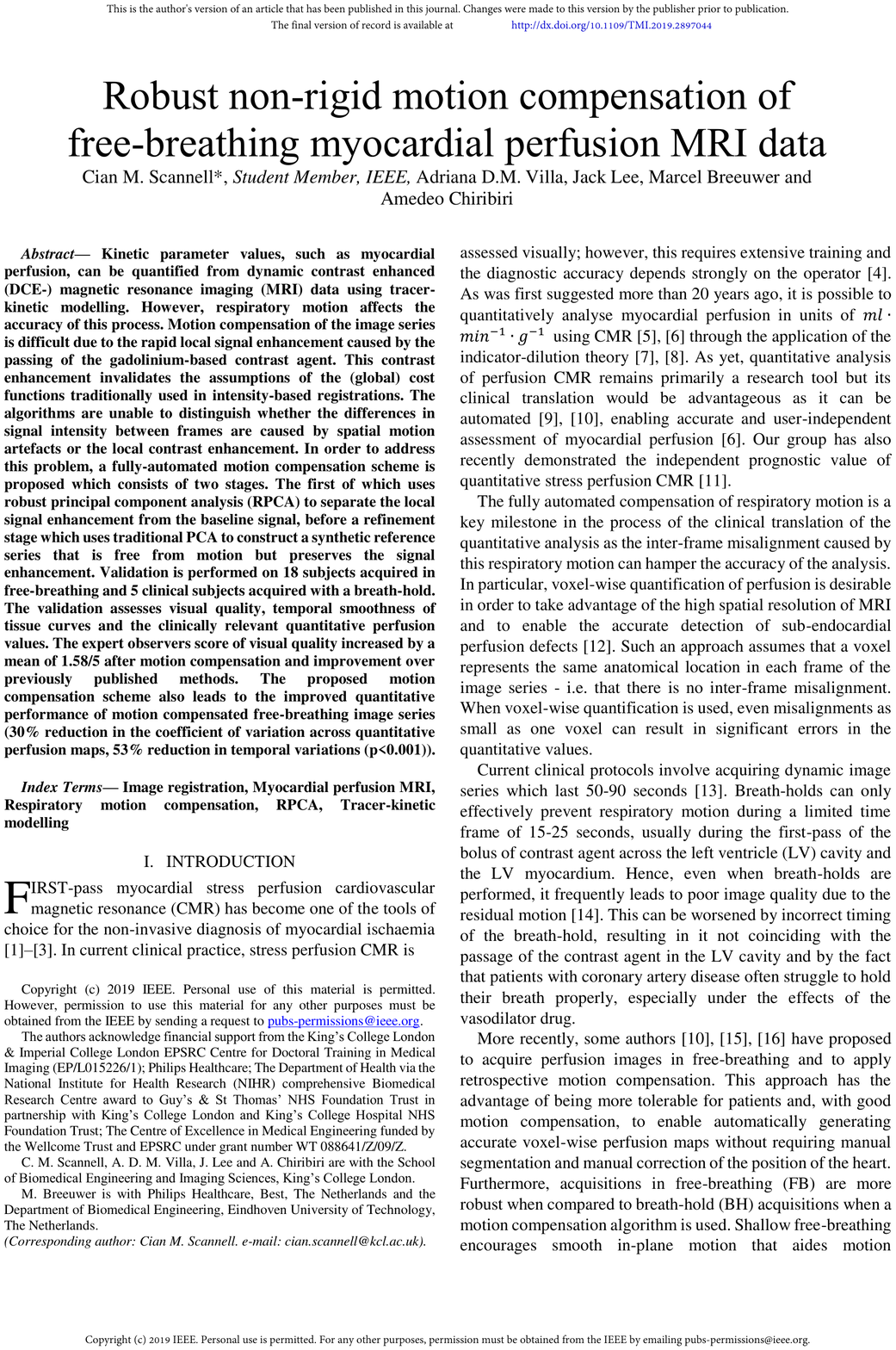}

\chapter[Automated image processing]{Deep learning‐based preprocessing for quantitative myocardial perfusion MRI}

\section{Preface} 

The field of image processing was revolutionised when a deep convolutional neural network, AlexNet \cite{Krizhevsky2012}, won the ImageNet Large Scale Visual Recognition Challenge by a large margin in 2012 \cite{Russakovsky2015}. This was the first demonstration that modern computer hardware, such as graphics processing units (GPUs), could be combined with large databases to train deep neural networks to successfully perform computer vision tasks.

Since this, deep learning has been widely adopted in the field of medical imaging and has become the de facto standard for many processing tasks \cite{Litjens2017}. Trends in cardiac MRI image analysis have followed a similar route where deep learning is now used for everything from reconstruction to detection and segmentation tasks to automating diagnostics and prognostics \cite{Leiner2019}.

\begin{figure}[h]
	\centering
	\includegraphics[width=\textwidth]{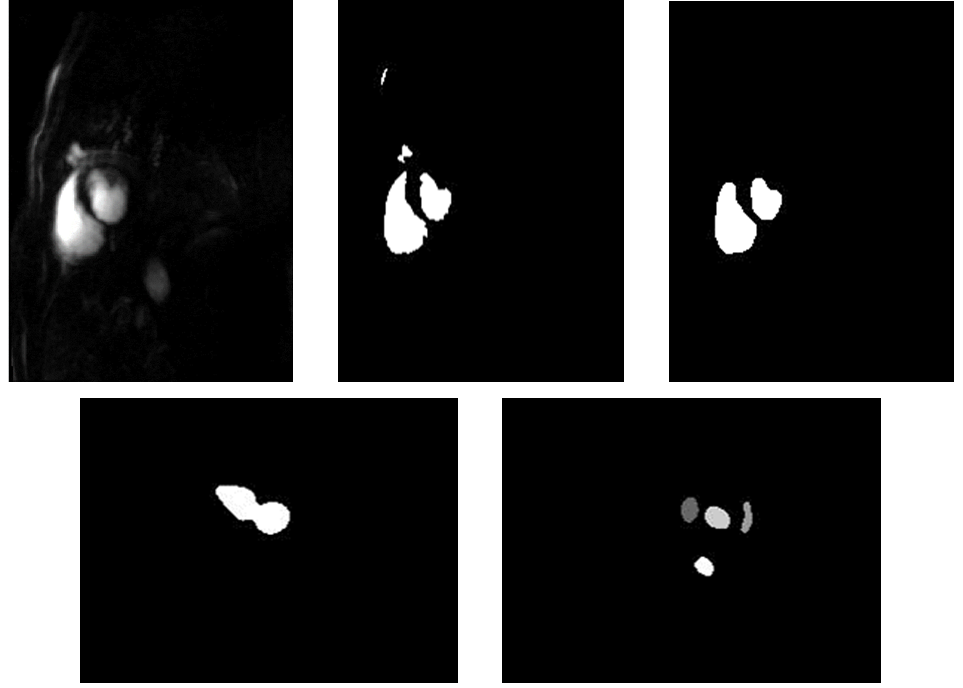}
	\caption[Failed bounding box detection.]{The top row shows the temporal standard deviation image, the thresholded standard deviations, and the result of the connected component analysis to identify the RV and LV (from left to right). The bottom row shows two failed cases. On the left, there only one component (a joint LV and RV) is identified and on the right four similar components are identified.}
	\label{fig:bb}
\end{figure}

The main advantage of deep learning is that the networks learn the features to be used to complete the task from the data rather than these being hand-engineered. Such hand-crafted features are biased by preconceived ideas of the programmer and thus tend to be brittle. An example of this, in the context of perfusion CMR, is the problem of bounding box detection which is a common first step in processing pipelines. Before the advent of deep learning, a popular approach for this task was based on thresholding temporal variances. The logic behind this was that the contrast flowing causes the areas of highest temporal variance to be the RV and LV cavity. However, this approach was shown to fail in 4/44 case \cite{Tautz2011a} and we found similarly high rates of failure in our implementation. An example of this algorithm with some of the common failure modes is shown in Figure~\ref{fig:bb}. For this reason, deep learning is used in this work for the image processing required to turn the raw image series into concentration curves for tracer-kinetic modelling.

\subsection{Supervised deep learning for image processing}

\subsubsection*{Supervised learning}
Supervised learning attempts to learn a model that takes input data and outputs labels \cite{Goodfellow2016}. That is, to learn the function $f$ which depends on parameters $\boldsymbol{\theta}$ that best maps the input space $\mathit{X}$ to the output space $\mathit{Y}$:
\begin{align} 
	f_{\boldsymbol{\theta}} : \mathit{X} \rightarrow \mathit{Y}.
\end{align} 

A training set of matched data and labels is required to find the best set of model parameters to map the input data to labels. The trained models can subsequently be applied to unseen test data to obtain a prediction of its label. This is opposed to unsupervised learning which attempts to learn structure in the data \cite{Goodfellow2016}.

\subsubsection*{Deep learning}
\begin{figure}[h]
	\centering
	\includegraphics[width=.8\textwidth]{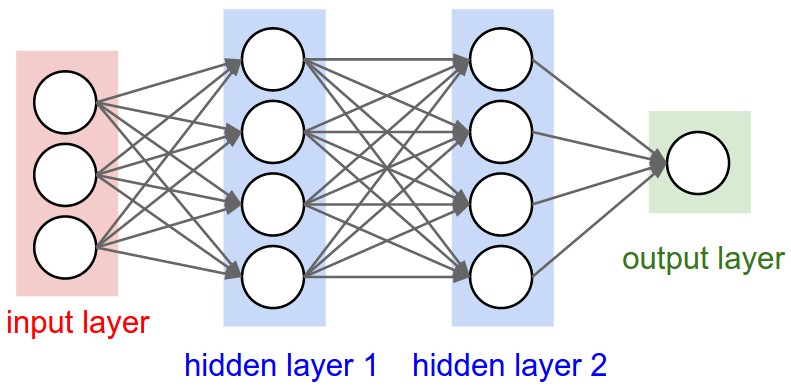}
	\caption[A fully-connected neural network]{A fully-connected neural network with two hidden layers, image taken from \url{https://cs231n.github.io/}.}
	\label{fig:fc_nn}
\end{figure}

Deep learning is the sub-field of machine learning based primarily on the use of neural networks \cite{Goodfellow2016}. Artificial neural networks are functions made up of layers of units or neurons. Each unit in a layer takes, as input, the output of the units in the previous layer. It then computes a weighted combination of these inputs and applies a non-linear activation function. The weights needed for the weighted combination are the parameters of the neural network and are optimised in order to solve the task at hand. The typical neural networks composes many internal layers and are thus described as deep. A simple neural network, with two hidden layers (layers in addition to the input and output layers), is illustrated in Figure~\ref{fig:fc_nn} As previously discussed, the early layers can be thought of as extracting features from the input data, while the later layers learn to combine these features to meet the objective. This yields a model function in the form:
\begin{align} \label{eqn:nn}
	f_{\boldsymbol{\theta}} (x) = f^L_{\boldsymbol{\theta_L}} (f^{L-1}_{\boldsymbol{\theta_{L-1}}} ( \cdots f^{0}_{\boldsymbol{\theta_{0}}} (x) \cdots  ))
\end{align}
where $f^i_{\boldsymbol{\theta_i}}$ is the $i$\textsuperscript{th} layer with parameters $\boldsymbol{\theta_i}$.

\subsubsection*{Convolutional neural networks}
\begin{figure}[h]
	\centering
	\begin{subfigure}{.5\textwidth}
		\centering
		\includegraphics[width=.9\linewidth]{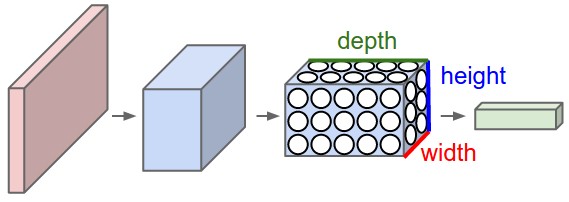}
		\label{fig:sub1}
	\end{subfigure}%
	\begin{subfigure}{.5\textwidth}
		\centering
		\includegraphics[width=.9\linewidth]{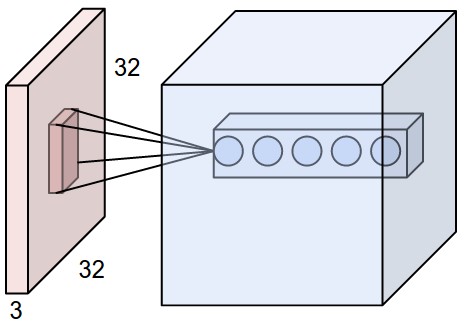}
		\label{fig:sub2}
	\end{subfigure}
	\caption[A convolutional neural network]{A convolutional neural network. This shows the layers of the CNN downsampling the input image and producing a 3D volume of activation (left), and a unit connected to a small receptive field in the previous layer (right), image taken from \url{https://cs231n.github.io/}}
	\label{fig:cnn}
\end{figure}

A network where all units are connected to all units in the previous layer is said to be fully-connected. This design is inefficient or infeasible for image processing because of the high dimensionality of the inputs (the dimensionality is equal to the number of pixels or voxels in the image). An alternative network design, more suitable for image processing, is the convolutional neural network (CNN), as visualised in Figure~\ref{fig:cnn}. A CNN combines inputs from only a small region of the previous layer (commonly referred to as the receptive field and designed to loosely resemble the visual cortex). This is implemented as a learnable kernel being convolved with the output of the previous layer to generate the input for the current layer. The convolutional kernel is designed to be significantly smaller than the size of the activation it is being convolved with. This enforces sparse interactions between layers and allows the extraction of low level features, such as edges, without considering the whole image. Parameter sharing is also used such that the same kernel is applied to all inputs to a layer. This aids the training process by greatly reducing the number of parameters to be learned. CNNs also encourage translational invariance in the predictions. Since a kernel slides over the whole input it will detect the same features regardless of their position. This is useful, for example in classification tasks, when it only matters if an object is in an image and not where it is \cite{Goodfellow2016}. 

\subsubsection{U-Net}
\begin{figure}[h]
	\centering
	\includegraphics[width=.8\textwidth]{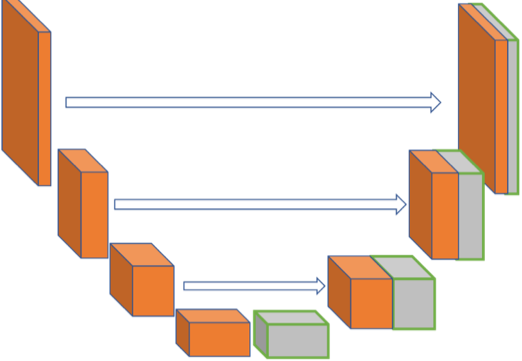}
	\caption[The U-Net model architecture]{The U-Net model architecture, image adapted from \url{https://www.nist.gov/}.}
	\label{fig:unet}
\end{figure}

The most commonly used network architecture for medical image processing is the U-Net \cite{Ronneberger2015}. This is a fully-convolutional network in that it only uses convolutional layers, and is illustrated in Figure~\ref{fig:unet}. The architecture has an encoder-decoder structure. The encoder downsamples the input image, using max-pooling, to create a low-dimensional embedding. The multi-step downsampling allows the learning of feature representations at different image scales. The decoder, takes the low-dimensional embedding and upsamples it to try predict the desired output, typically a segmentation map. The network also utilises skip connections that concatenate the activations of the encoder to the corresponding resolution level of the decoder. This is thought to better allow the recovery of the fine-grained details in the prediction.

\subsubsection*{Training neural networks}
Neural networks can be trained (the optimal parameters estimated) using first-order stochastic optimisation algorithms to minimise a loss function $\mathcal{L}$ between the estimated network output $\hat{y}$ and the training labels $y$. The most simple approach to this is stochastic gradient descent: 
\begin{align}
	\boldsymbol{\theta} \leftarrow \boldsymbol{\theta} - \alpha \nabla_{\boldsymbol{\theta}} \mathcal{L}(y, \hat{y})
\end{align}
where $\alpha$ is the learning rate that controls the size of the parameter updates made in the optimisation. The algorithm is referred to as stochastic as it does not consider the whole dataset at each iteration, for computational efficiency. Parameter updates are instead made using the gradients computed on a small sub-sample of the training data, known as a mini-batch. Due to the layer-wise nature of neural networks, as seen in Equation~\ref{eqn:nn}, the gradient of the loss with respect to the parameters of a layer can be computed using the chain rule, in a process is known as backpropagation.

A more advanced (and recently more popular) gradient descent update rule is the ADAM optimiser \cite{Kingma2014} which uses adaptive learning rates and momentum to achieve better convergence properties \cite{Goodfellow2016}.

\section{Journal article} \label{section:dl_paper}
The following text is reproduced as published \cite{Scannell2020}:

\vspace{3mm}

\noindent \textit{\textbf{Scannell, C.M.}, Veta, M., Villa, A.D.M., Sammut, E., Lee, J., Breeuwer, M. \& Chiribiri, A. Deep‐Learning‐Based Preprocessing for Quantitative Myocardial Perfusion MRI. J. Magn. Reson. Imaging 51, 1689–1696 (2020).}

\includepdf[pages=-,pagecommand={},width=1.2\textwidth]{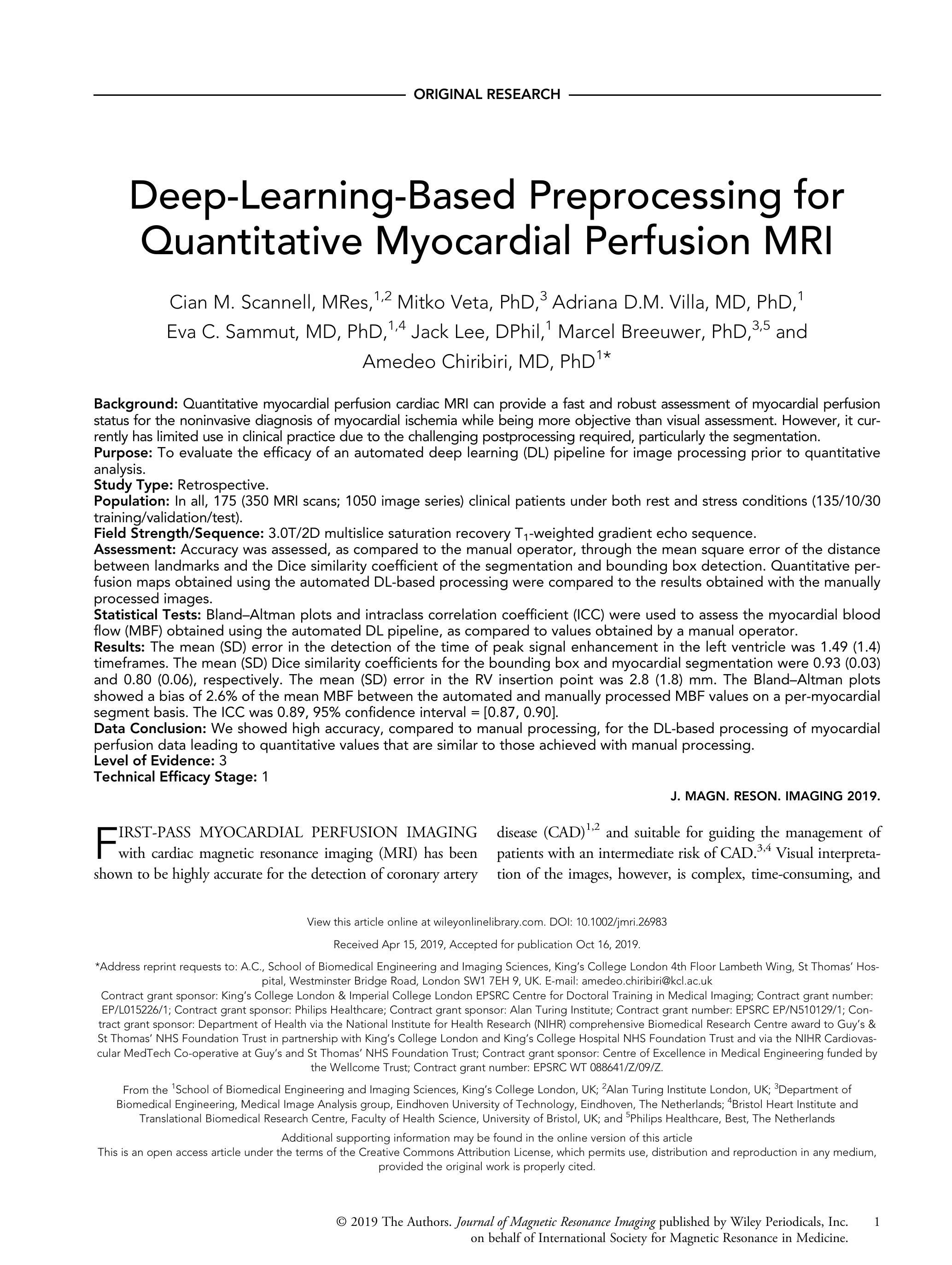}

\section{Supplementary material}
\begin{table}[h]
	\centering
	\begin{tabular}{ll}
		\hline \hline  
		& N = 175               \\ \hline  
		Male gender                    & 136 (78\%)            \\
		Age (years)                    & 64.3 $\pm$ 10.3       \\
		Hypertension                   & 86 (49\%)             \\
		Diabetes                       & 34 (19\%)             \\
		Hypercholesterolemia           & 78 (45\%)             \\
		Current / previous smoker      & 24 (14\%) / 18 (10\%) \\
		CAD status (visual assessment) & -                     \\
		- 1 vessel                     & 43 (25\%)             \\
		- 2 vessels                    & 28 (16\%)             \\
		- 3 vessels                    & 31 (18\%)             \\
		\hline \hline 
	\end{tabular}
	\caption{Demographic characteristics of the population.}
\end{table}

\begin{table}[h]
	\centering
	\begin{tabular}{llll}
		\hline \hline
		Layer & Input size & Convolutional kernel & Number of filters \\ \hline
		1     & 256 x 256  & 3 x 3                & 8                 \\
		2     & 128 x 128  & 3 x 3                & 16                \\
		3     & 64 x 64    & 3 x 3                & 32                \\
		4     & 32 x 32    & 3 x 3                & 64                \\
		FC 1  & 8192       & -                    & -                 \\
		FC 2  & 512        & -                    & -                \\
		\hline \hline
	\end{tabular}
	\caption[The architecture used for the peak LV enhancement frame detection and bounding box detection.]{The architecture used for the peak LV enhancement frame detection and bounding box detection. FC 1 and FC 2 are the fully connected layers. Each convolutional layer involves convolutional with the filter with a stride length of 2 followed by batch normalisation and ReLU activation. Max-pooling is performed after every convolutional layer.}
\end{table}

\begin{table}[h]
	\centering
	\begin{tabular}{llll}
		\hline \hline
		Layer & Input size & Convolutional kernel & Number of filters \\ \hline
		1-3   & 96 x 96    & 3 x 3                & 16                \\
		4-6   & 48 x 48    & 3 x 3                & 32                \\
		6-9   & 24 x 24    & 3 x 3                & 64                \\
		9-12  & 12 x 12    & 3 x 3                & 128               \\
		12-15 & 6 x 6      & 3 x 3                & 256               \\
		15-18 & 12 x 12    & 3 x 3                & 128               \\
		18-21 & 24 x 24    & 3 x 3                & 64                \\
		21-24 & 48 x 48    & 3 x 3                & 32                \\
		24-27 & 96 x 96    & 3 x 3                & 16               \\
		\hline \hline
	\end{tabular}
	\caption[The U-Net architecture used for the myocardial segmentation and the RV insertion point detection.]{The U-Net architecture used for the myocardial segmentation and the RV insertion point detection. On the downward trajectory (layers 1-12) each layer involves convolutional with the filter with a stride length of 2 followed by batch normalisation and ReLU activation. Max-pooling is performed after every third convolutional layer to down-sample the image dimensions. On the upward trajectory (layers 16-27), each layer still involves convolutional with the filter with a stride length of 2 followed by batch normalisation and ReLU activation with transposed convolutions being performed before every third layer to up-sample the image dimensions. Features are concatenated across the layers of the same scale, as previously described.}
\end{table}

\begin{figure}[h]
	\centering
	\includegraphics[width=.8\textwidth]{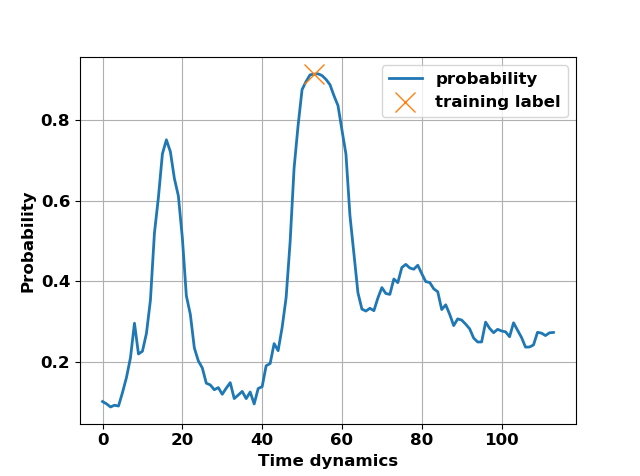}
	\caption[The predicted probability of being the peak LV enhancement frame over time.]{The probability of being the peak LV enhancement frame over time. An increased probability is seen as contrast arrives in the LV, reaching a peak with the peak signal frame (marked with an X) and then reducing as the contrast washes out. The network also assigns relatively high probabilities to the time frames corresponding to the pre-bolus injection.}
	\label{fig:probsLV}
\end{figure}

\begin{figure}[h]
	\centering
	\includegraphics[width=.9\textwidth]{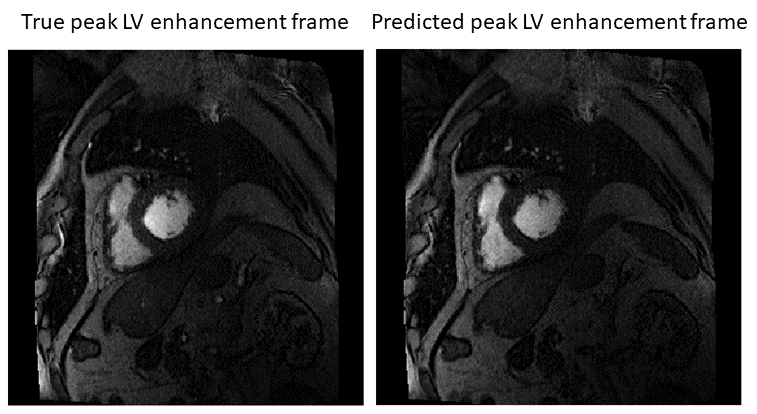}
	\caption[The worst test prediction for peak LV enhancement time frame.]{A comparison of the automated and manually chosen peak LV enhancement time frame for the patient with the largest error (3 beats) in the test set. The 2 frames are virtually indistinguishable on visual assessment.}
	\label{fig:worst_LV_time}
\end{figure}

\begin{figure}[h]
	\centering
	\includegraphics[width=\textwidth]{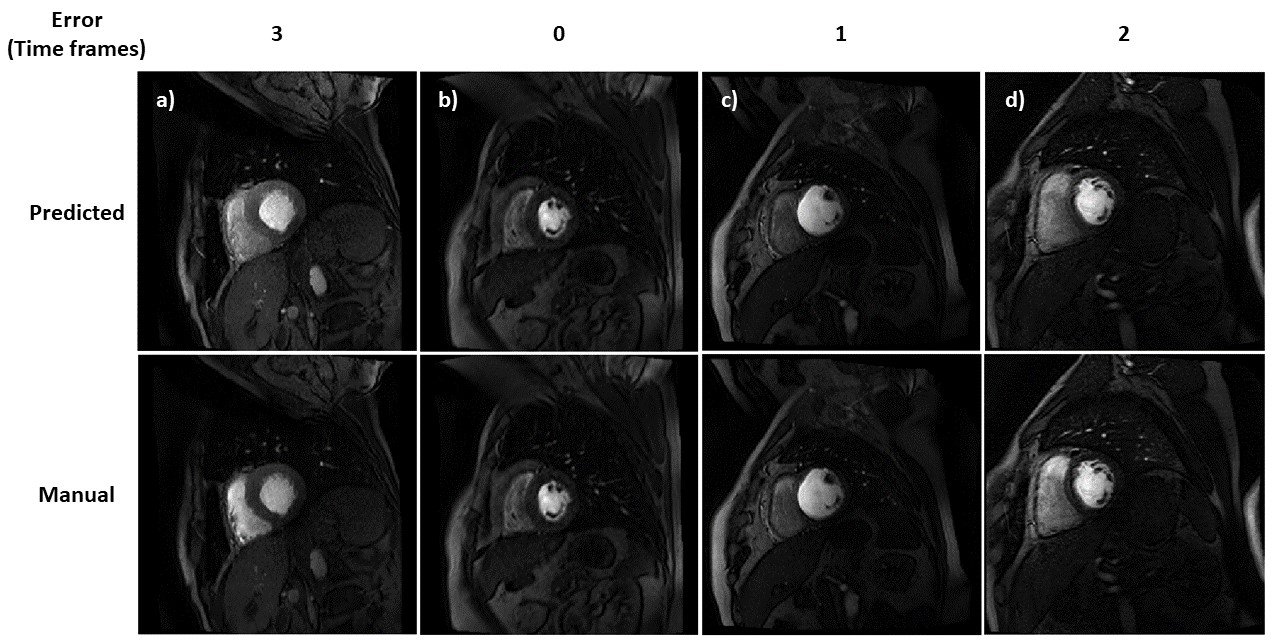}
	\caption[Automated versus manually chosen peak LV enhancement time frame.]{A comparison of the automated and manually chosen peak LV enhancement time frame from a representative set of patients from the test set. The error is in terms of the number of time frames.}
	\label{fig:LV_time_preds}
\end{figure}

\begin{figure}[h]
	\centering
	\includegraphics[width=\textwidth]{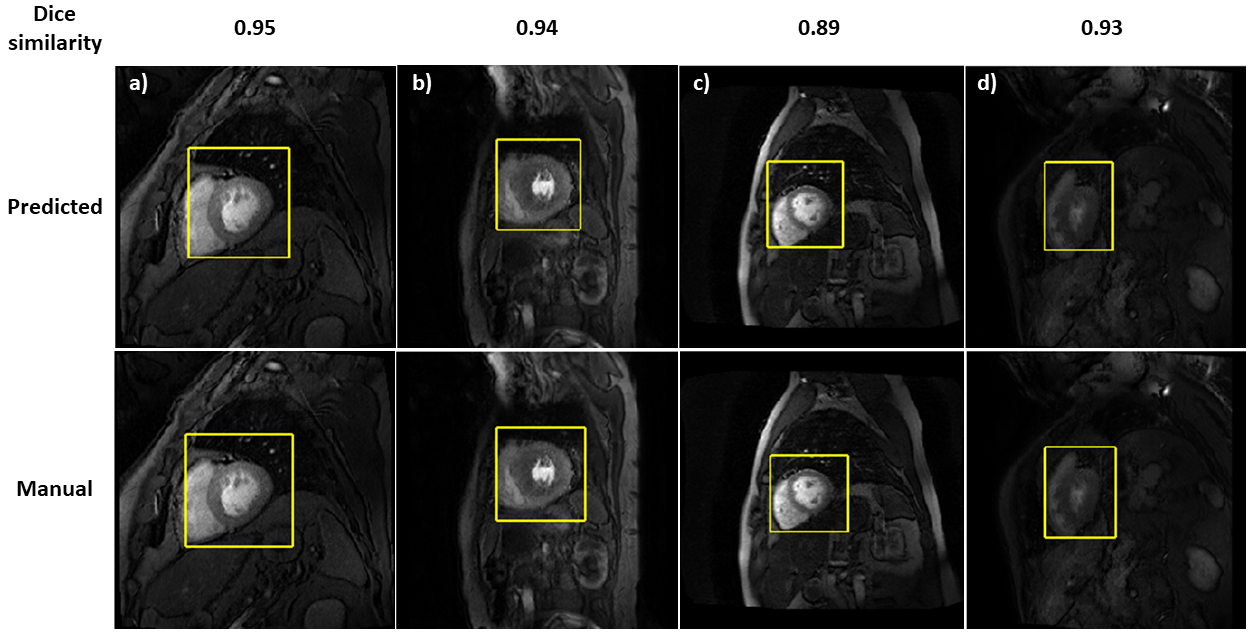}
	\caption[Automated versus manually determined bounding boxes.]{A representative set of patients from the test set with a comparison between the automatically and manually determined bounding boxes.}
	\label{fig:LV_time_crops}
\end{figure}

\begin{figure}[h]
	\centering
	\includegraphics[width=\textwidth]{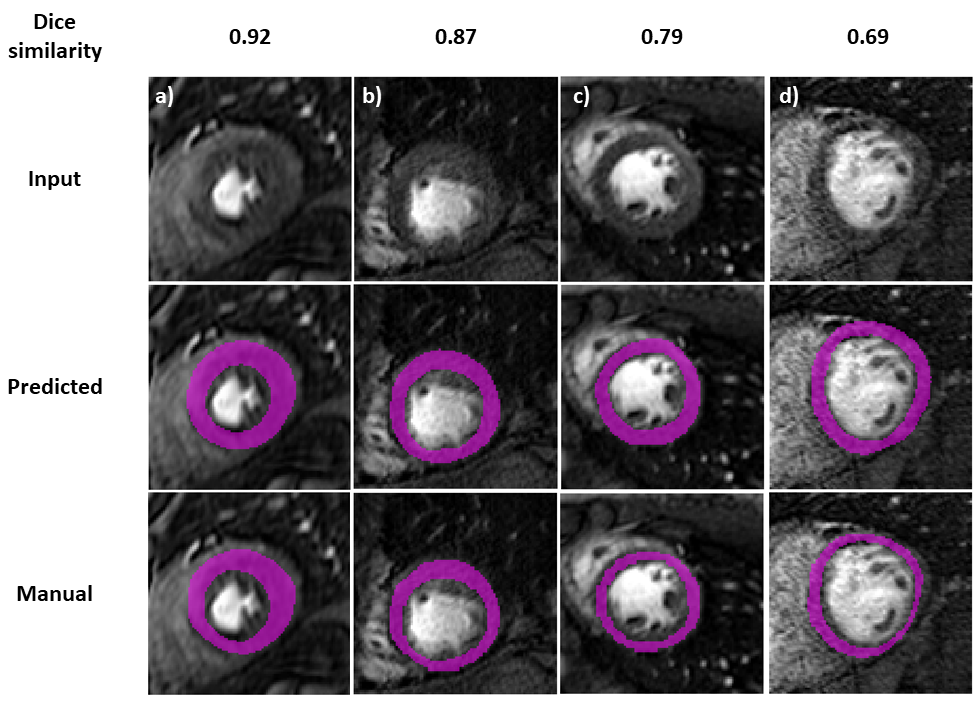}
	\caption[Automated versus manual segmentations.]{A representative set of patients from the test set with a comparison between the automated segmentation and the manually defined segmentations.}
	\label{fig:seg_preds}
\end{figure}

\begin{figure}[h]
	\centering
	\includegraphics[width=\textwidth]{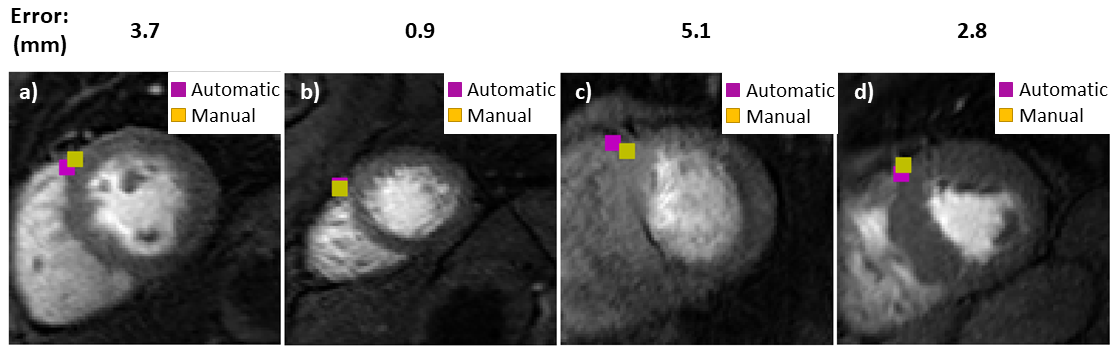}
	\caption[Automated versus manually determined RV insertion points.]{A representative set of patients from the test set with a comparison between the automatically detected RV insertion point and the manually defined RV insertion point (error is in mm).}
	\label{fig:RV_preds}
\end{figure}

\begin{figure}[h]
	\centering
	\includegraphics[width=\textwidth]{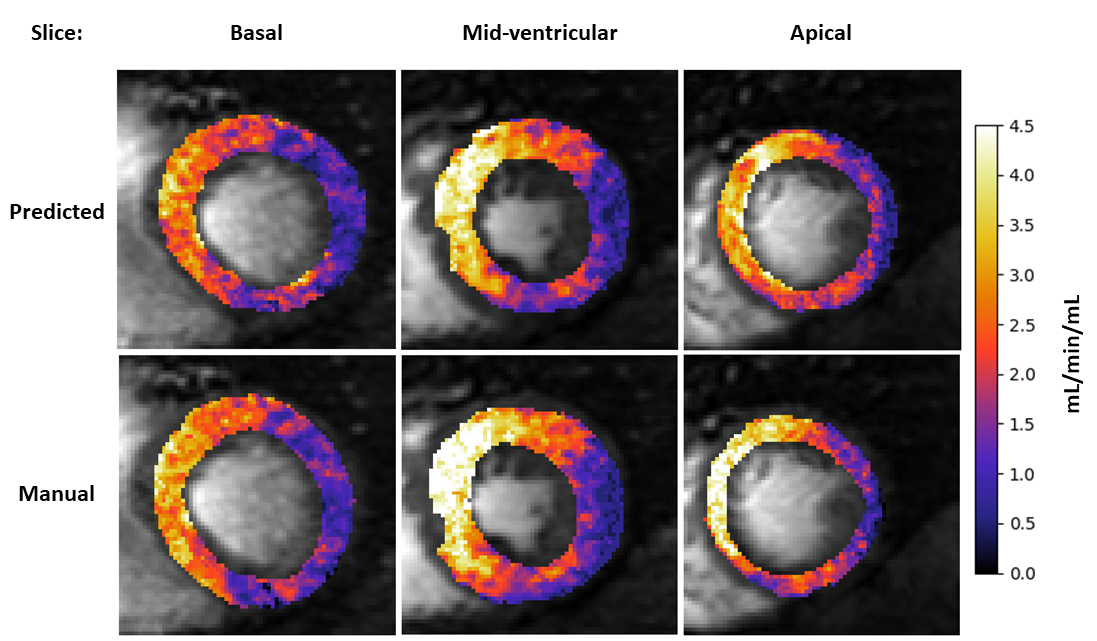}
	\caption[Automated versus manual MBF maps.]{A representative patient from the test set with a comparison between the automatically processed quantitative perfusion maps and the manually processed quantitative perfusion maps. Coronary angiography showed the patient has CAD with a lesion in the proximal left circumflex coronary artery.}
	\label{fig:quants_pred}
\end{figure}

\begin{figure}[h]
	\centering
	\includegraphics[width=\textwidth]{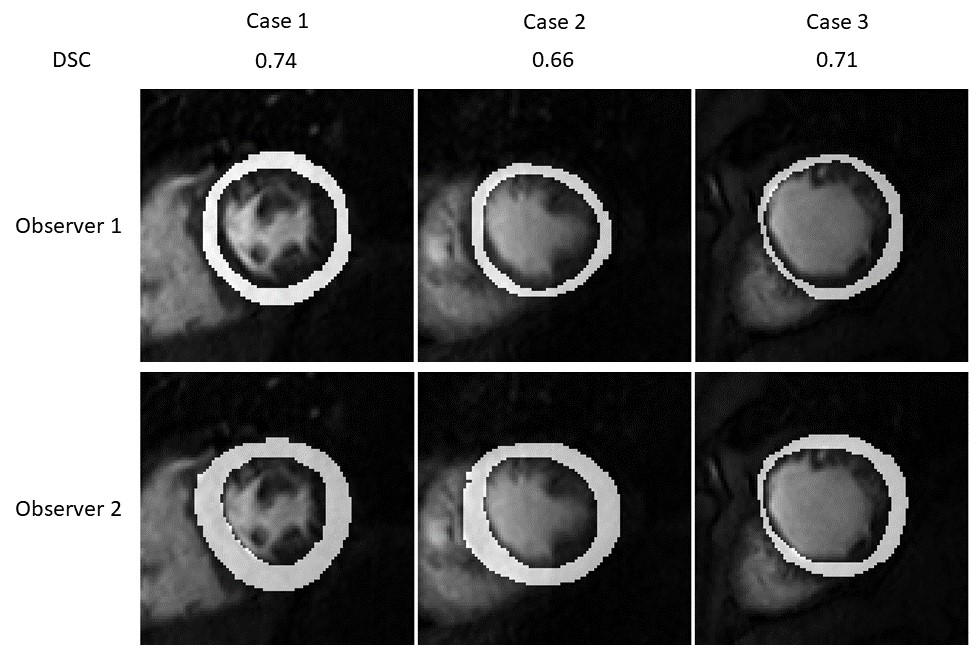}
	\caption[Inter-observer variability of myocardial segmentation.]{A comparison between the manual segmentations obtained from two different observers.}
	\label{fig:inter_ob}
\end{figure}

\chapter[Bayesian kinetic parameter inference]{Hierarchical Bayesian myocardial perfusion quantification}

\section{Preface} 
As discussed in Section~\ref{section:tkmodel}, the quantification of myocardial perfusion is an inverse problem. Given an observed AIF and myocardial tissue curve, the problem is to find the kinetic parameters such that the model best matches the observed data. 

The question of whether or not the parameters are recoverable from the data is known as identifiablity. Romain et al. \cite{Romain2017} showed that tracer-kinetic models of the form considered in this thesis are structurally identifiable. This is a strictly theoretical proof that states that the parameters are exactly recoverable in ideal, noise-free, infinite temporal resolution case because the mapping from the parameters to the residue function is a bijection.

The more pressing concern for the use of such models in real-world settings is the practical identifiablity. Practical identifiablity assesses the feasibility of recovering the correct parameters given the limited, imperfect measurements available. This has long been called into question for myocardial perfusion modelling and DCE-MRI in general. In 2002, Buckley \cite{Buckley2002} reported on the vast amount of possible parameter combinations that are indistinguishable at the noise level present in the data: "the issue of parameter uniqueness is further complicated by the surfeit of possible parameter combinations" and remarked that this introduces significant uncertainty in the estimates. This uncertainty has been similarly noted by an array of other authors \cite{Jerosch-Herold1999, Broadbent2013, Schwab2015} and specific to myocardial perfusion CMR, Likhite et al. \cite{Likhite2017} reported the same phenomenon:
"A shortcoming with complex pharmacokinetic models is that identical tissue curves can be generated using a single arterial input function and multiple sets of perfusion model parameters".

However, it is of course possible to get the correct parameters, as is evidenced by the many successful studies conducted in the field
\cite{Broadbent2016,Papanastasiou2016,Sammut2017,Biglands2018,Knott2020}. It is a question of the trade-off between the amount of information in the data and the complexity of the model to be estimated. For example, the amount of information in the data can be increased by fitting the model on an AHA segment level rather than a pixel-wise level \cite{Broadbent2016,Biglands2018}. On the other hand, the complexity of the model can be reduced by considering models with less parameters, for example, the Fermi model \cite{Sammut2017}.

A relatively unexplored approach to increasing the amount of information available in the data is through the use of prior knowledge. This approach is investigated in this thesis, in a Bayesian inference framework.

\subsection{Bayesian kinetic parameter estimation.}

If we define $C_{\boldmath{\theta}}(t) = F_p \cdot R(t;\boldsymbol{\theta}) \ast C_{AIF}(t)$ to be the output of the tracer-kinetic model for a given set of parameters $\boldmath{\theta}$, $T$ to be the total number of time points, and $\mathbf{y} = (y_1, y_2, \ldots, y_T)$ to be the observed myocardial concentrations at times $\mathbf{t} = (t_1, t_2, \ldots, t_T)$, then the standard non-linear least squares (NLLS) parameter estimate is given as: 

\begin{align} \label{eqn:nlls}
	\hat{\boldmath{\theta}} = \argmin_{\boldmath{\theta}} \frac{1}{n} \sum_{i=1}^{T} (y_i - C_{\boldmath{\theta}}(t_i))^2.
\end{align}
It can be seen that $\argmin_x x = \argmax_x \exp(-x)$ and since constant scaling factors do not affect the locations of minima or maxima, Equation~\ref{eqn:nlls} is equivalent to:
\begin{align}
\hat{\boldmath{\theta}} & = \argmax_{\boldmath{\theta}} \frac{1}{(2\pi)^{n/2}\sigma^n } \exp \bigg( \frac{- \sum_{i=1}^{T} (y_i - C_{\boldmath{\theta}}(t_i))^2 }{2\sigma^2} \bigg) \\
& = \argmax_{\boldmath{\theta}} \mathcal{N}(C_{\boldmath{\theta}}(\mathbf{t}), \sigma^2).
\end{align}
where $\mathcal{N}(\mu,\sigma^2)$ is the Gaussian distribution with mean $\mu$ and variance $\sigma^2$. The term being maximised is known as the likelihood function, and is also written as $P(\mathbf{y}|\boldmath{\theta})$, this can be interpreted as the likelihood of observing the data that was observed, conditioned on the parameters. It is therefore seen that the NLLS solution is the maximum likelihood parameter estimate, under the assumption of Gaussian noise:
\begin{align} \label{eqn:nlls}
\hat{\boldmath{\theta}} = \argmax_{\boldmath{\theta}} P(\mathbf{y}|\boldmath{\theta})
\end{align}

As previously discussed, the least squares solution can possess local optima and the global minima can be hard to find. This is a limitation as the method returns a single point estimate that must be assumed to be the true value. An alternative approach, Bayesian inference, is to assume that there is a distribution of possible parameter values and to try to compute this distribution. Bayes theorem allows the posterior distribution of the parameters to be written in terms of the likelihood function as:
\begin{align}
	P(\boldmath{\theta}|\mathbf{y}) &= \frac{P(\mathbf{y}|\boldmath{\theta}) \cdot P(\boldmath{\theta})}{P(\mathbf{y})}  
\end{align}
where $P(\boldmath{\theta})$ is the prior probability of the data and $p(\mathbf{y}) = \int_{\boldmath{\theta}} P(\mathbf{y}|\boldmath{\theta}) \cdot P(\boldmath{\theta}) d\boldmath{\theta}$ is the probability of the data. From the posterior distribution, the expected value of the parameters can be computed and also the standard deviation of the distribution gives a measure of the uncertainty of the parameter estimates. The difficultly is that the posterior distribution is not, in general, analytically tractable as it involves integrals which can not be computed and thus, numerical solutions must be considered.

\subsection{Metropolis-Hastings}
The Metropolis-Hastings algorithm is commonly used to numerically approximate posterior distributions. There are some definitions required to introduce the Metropolis-Hastings algorithm: a Markov chain is a random process, with  discrete time steps, that has the Markov property. The Markov property states that the conditional probability of a future state of the random process depends only on the current state and not on any of the past states. The limiting distribution of a Markov chain is the distribution that it converges to asymptotically. Finally, a Monte Carlo method is simply a method that uses random samples to solve a problem.

Then, the Metropolis-Hastings algorithm is a Markov chain Monte Carlo (MCMC) method to approximate a distribution by simulating samples from it. It aims to construct a Markov chain with limiting distribution that is the distribution being approximated. It does this by taking a random walk through the distribution space, where the steps are proposed by a proposal distribution. If the posterior probability of the proposed position is greater than the posterior probability of the current position, the proposed step is accepted. If not, the proposed step can also be accepted with a probability proportional to the reduction in posterior probability it would cause. With the notation that samples $\boldmath{\theta}$ are being sampled using a proposal distribution $q$ from the target distribution $\pi(\cdot)$, the Metropolis-Hastings algorithm is given by Algorithm~\ref{algo:mh}.\\

\vspace{3mm}
\begin{algorithm}[H] \label{algo:mh}
	\SetKwInOut{Input}{Input}
	\SetKwInOut{Output}{Output}
	\SetKwInOut{Initialise}{Initialise}
	
	\Initialise{$\boldmath{\theta^0}$}
	
	\For{iteration $i = 1,2,\ldots,N $}{
		$\boldmath{\theta^{cand}} \sim q(\boldmath{\theta^{i}} |\boldmath{\theta^{i-1}} )$\;
		$\alpha = \min\{1, \frac{q(\theta^{i-1}|\theta^{cand}) \pi(\theta^{cand}))}{q(\theta^{cand}|\theta^{i-1})\pi(\theta^{i-1})} \}$\;
		$u \sim \textnormal{Uniform}[0,1]$\;
		\eIf{$u < \alpha$}{
			$\boldmath{\theta^{i}} \leftarrow \boldmath{\theta^{cand}}$\;
		}
		{ 
			$\boldmath{\theta^{i}} \leftarrow \boldmath{\theta^{i-1}}$\;
		}

	}
	\Output{ $(\boldmath{\theta^{0}},\boldmath{\theta^{1}},\ldots,\boldmath{\theta^{N}})$}
	\caption{Metropolis-Hastings}
\end{algorithm}
\vspace{3mm}
\noindent It is common to use a symmetric proposal distribution such that $q(a|b) = q(b|a)$ such that:
\begin{align}
	 \frac{q(\theta^{i-1}|\theta^{cand}) \pi(\theta^{cand}))}{q(\theta^{cand}|\theta^{i-1})\pi(\theta^{i-1})}  =  \frac{\pi(\theta^{cand}))}{\pi(\theta^{i-1})}. 
\end{align}
It should also be noted that $\pi$ is only required up to a constant scaling factor (as this would cancel in the numerator and denominator) so when approximating the posterior distribution for parameter inference, it is sufficient to use $P(\mathbf{y}|\boldmath{\theta}) \cdot P(\boldmath{\theta})$ as:
\begin{align}
P(\boldmath{\theta}|\mathbf{y}) \propto P(\mathbf{y}|\boldmath{\theta}) \cdot P(\boldmath{\theta}).
\end{align}

Metropolis-Hastings is the approach used for the parameter inference in this thesis. However, there are more efficient sampling approaches being developed, such as Hamiltonian Monte Carlo sampling \cite{Betancourt2017a}, which use gradient information to inform proposals, rather than random walks. This encourages movement towards areas of high probability and thus more efficient samples.

\section{Journal article} \label{section:bayes_paper}
The following text is reproduced as published \cite{Scannell2020a}:

\vspace{3mm}

\noindent \textit{\textbf{Scannell, C.M.}, Chiribiri, A., Villa, A.D.M., Breeuwer, M. \& Lee, J. Hierarchical Bayesian myocardial perfusion quantification. Med. Image Anal. 60, 101611 (2020).}

\includepdf[pages=-,pagecommand={},width=1.2\textwidth]{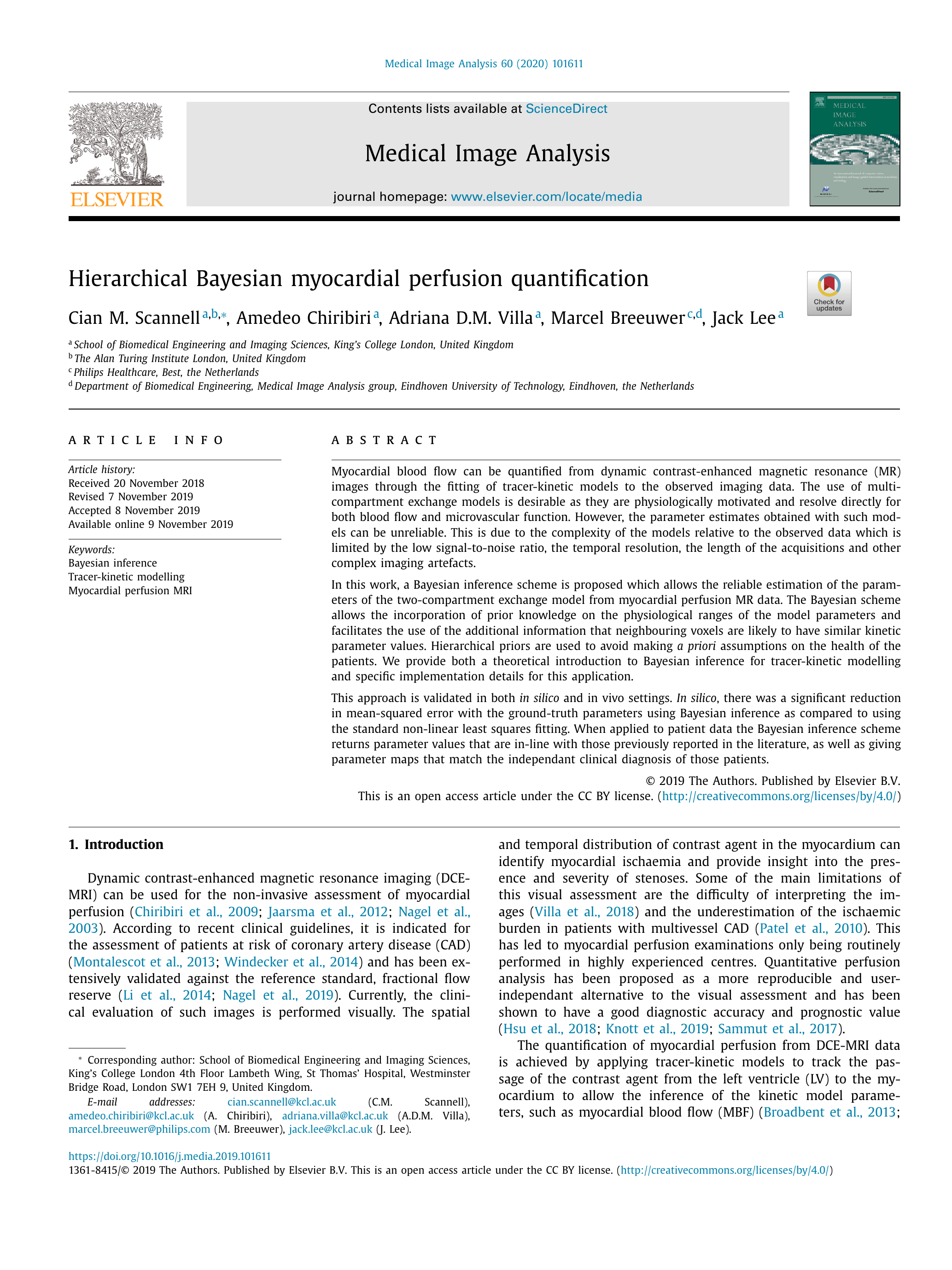}

\section{Correction}
In the appendix of Section~\ref{section:bayes_paper}, it is stated that:
\begin{align}
R_F(t, \boldmath{\theta}) = A \exp(\alpha t) + (1-A) \exp(\beta t).
\end{align} 
This should read as: 
\begin{align}
R_F(t, \boldmath{\theta}) = F_p \cdot (A \exp(\alpha t) + (1-A) \exp(\beta t)). 
\end{align}
\chapter{Preliminary evaluation}

\section{Introduction}
As discussed, stress perfusion CMR is now an established technique for the assessment of patients with CAD \cite{Greenwood2012,Schwitter2013a} and it is playing an expanding role in the assessment of patients with angina and coronary microvascular dysfunction (MVD) \cite{Rahman2019}. Although a high diagnostic accuracy and prognostic value is found with visual assessment at experienced centres, quantification is required to make this a widespread clinical reality. Therefore, this chapter presents a preliminary, proof of principle evaluation of the automated pipeline for stress perfusion quantification that was presented in Chapters 4, 5, and 6. In particular, the ability of stress perfusion CMR to detect myocardial ischaemia is demonstrated, both in patients with obstructive epicardial CAD and MVD. This paves the way for a more thorough validation study and the subsequent clinical adoption of quantitative stress perfusion CMR. 

\section{Material and methods}
\subsection{Patients}

In total, 47 patients were included in this study. Of these, 36 patients with known or suspected CAD referred on clinical grounds for a stress perfusion CMR, who had a subsequent invasive coronary angiography, were retrospectively included. A further 11 patients who had MVD diagnosed by exclusion, demonstrable ischaemia with angiographically smooth coronary arteries, were also included. All patients had invasive coronary angiography within 3 months and the cut-off point for coronary lumen stenosis was 70\% for epicardial vessels. All invasive angiographic images have been reviewed by consensus of expert operators.

\subsection{Image acquisition}
All examinations were performed on a 3T system (Achieva TX, Philips Healthcare, Best, The Netherlands) using a 32-channel cardiac phased array receiver coil. Perfusion images were acquired in 3 LV short-axis slices (apical, mid-cavity, and basal) at mid-expiration with a saturation-recovery gradient echo method (repetition time 3.0 ms, echo time 1.0 ms, flip angle 15°, saturation-recovery delay 120 ms, sensitivity encoding (SENSE) parallel imaging, representative spatial resolution 1.5x1.5x10 mm\textsuperscript{3}). Stress images were acquired during adenosine-induced hyperaemia (140$\mu$g/kg/min). 0.075 mmol/kg of bodyweight gadolinium (Gd) extracellular contrast agent (gadobutrol, Gadovist, Bayer, Germany) was injected at 4 mL/s followed by a 20-mL saline flush for each perfusion acquisition. Each bolus of gadobutrol was preceded by a diluted pre-bolus with 10\% of the dose to allow quantification of perfusion, according to published methods \cite{Ishida2011}.

\subsection{Image processing}
The perfusion images were corrected for respiratory motion (Chapter 4) and processed fully-automatically using our deep learning-based processing pipeline (Chapter 5). Pixel-wise time signal intensity curves were extracted from the myocardial mask and signal intensity curves were subsequently split into the time intervals corresponding to the pre bolus injection and the main bolus injection for quantification, using Bayesian inference (Chapter 6). The pixel-wise MBF estimates were assigned to standard AHA segments using the automatically computed RV insertion points. AHA segments are attributed to coronary perfusion territories as described by Cerqueira et al. \cite{Cerqueira2002}. Each AHA segment is further sub-divided into an endocardial and epicardial layer to yield 32 segments. For the purpose of detecting CAD, Lockie et al. \cite{Lockie2010} validated the use of the mean value of the two lowest MBF values in a perfusion territory to represent that vessel. This work uses the mean of the four lowest MBF values as our segments are half the size. 

\subsection{Quantitative analysis}
A series of comparisons are performed. The first of which is to assess the ability of quantitative stress CMR to detect ischaemia and thus compares the patients with no significant CAD (CAD-) against the groups of patients with ischaemia, that is those with significant CAD (CAD+) or MVD. The second assessment is the ability to distinguish between significant CAD and no significant CAD and compares the CAD+ and CAD-. This is done on both a per-patient and per-coronary vessel level. The final assessment is the ability to detect MVD in the setting of patients with anatomically smooth coronary arteries and thus compares CAD- against MVD.

\subsection{Statistical analysis}
The data are presented as mean $\pm$ standard deviation (SD) and compared using two-sided Student \textit{t}-tests. p values < 0.05 were considered statistically significant. ROC curve analysis was used to report the diagnostic accuracy of the proposed approach. The data was analysed on both the per-patient and per-coronary vessel level. All statistical analysis was performed using SciPy and Pingouin \cite{Virtanen2020,Vallat2018a}.

\section{Results}
The baseline patient characteristics are summarised in Table~\ref{table:demos}. 17 patients had no significant coronary lesions or MVD. In 19 patients with CAD, there was 38 vessels with significant lesions and 11 patients had MVD. The findings of the angiography data are shown in Table~\ref{table:angios}.
\begin{table}[] 
	\centering
	\begin{tabular}{l|c}
		& \multicolumn{1}{c}{n = 47}   \\ \hline \hline
		Age (years)             & 59.66 $\pm$ 8.03             \\
		Male                    & 26 (55)                      \\
		BMI                     & 28.11 $\pm$ 3.89             \\
		Hypertension            & 22 (47)                      \\
		Hypercholesterolemia    & 25 (53)                      \\
		Diabetes                & 14 (30)                      \\
		Smoker                  & 8 (17)                       \\
		Ex-smoker               & 14 (30)                      \\
		Previous PCI            & 12 (26)                      \\
		Previous CABG           & 0 (0)                        \\ \hline
		\multicolumn{2}{l}{Values are n (\%) or mean $\pm$ SD}
	\end{tabular} 
\caption[Baseline demographics of the patient cohort.]{Baseline demographics of the patient cohort. PCI, percutaneous coronary intervention; CABG, coronary artery bypass grafting.} \label{table:demos}
\end{table}
\begin{table}[] 
	\centering
	\begin{tabular}{l|c}
		& \multicolumn{1}{c}{n = 47}   \\ \hline \hline
		No significant CAD           & 17 (36)             \\
		1-vessel disease                    & 7 (15)                      \\
		2-vessel disease                     & 2 (4)             \\
		3-vessel disease            & 10 (21)                      \\ 
		MVD           & 11 (23)                      \\ \hline
		\multicolumn{2}{c}{Values are n (\%)}
	\end{tabular} 
	\caption{Angiographic findings} \label{table:angios}
\end{table}

\begin{figure}[h!]
	\centering
	\includegraphics[width=.9\textwidth]{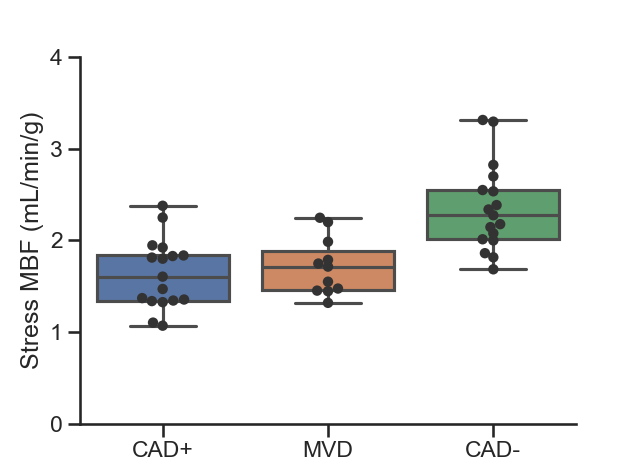}
	\caption[The distributions of global MBF values.]{The distributions of global MBF values in the different patient groups.}
	\label{fig:cad_mvd_mbf}
\end{figure}

45/47 patients were successfully analysed with two having failed contrast injections. The mean MBF value over all patients was $1.93 \pm 0.52$ \si{\milli\litre\per\min\per\gram}. The other microcirculatory parameters estimated were $v_b = 0.08 \pm 0.04$, $v_e = 0.18 \pm 0.08$, and $PS = 0.65 \pm 0.31$ \si{\milli\litre\per\min\per\gram}. The distribution of global MBF values for the three patient groups (CAD+, MVD, and CAD-) is shown in Figure~\ref{fig:cad_mvd_mbf}. The mean MBF in CAD- patients was $2.35 \pm 0.46$ \si{\milli\litre\per\min\per\gram}, in MVD patients was $1.72 \pm 0.3$ \si{\milli\litre\per\min\per\gram}, and in CAD+ patients was $1.67 \pm 0.36$ \si{\milli\litre\per\min\per\gram}. The MBF was significantly reduced in both the CAD+ and MVD groups as compared to the CAD- group (both p < 0.01), indicating the presence of significant ischaemia. There was no statistical difference in the mean MBF between the MVD and CAD+ patients ($p=0.53$).  

\begin{figure}[h!]
	\centering
	\includegraphics[width=.9\textwidth]{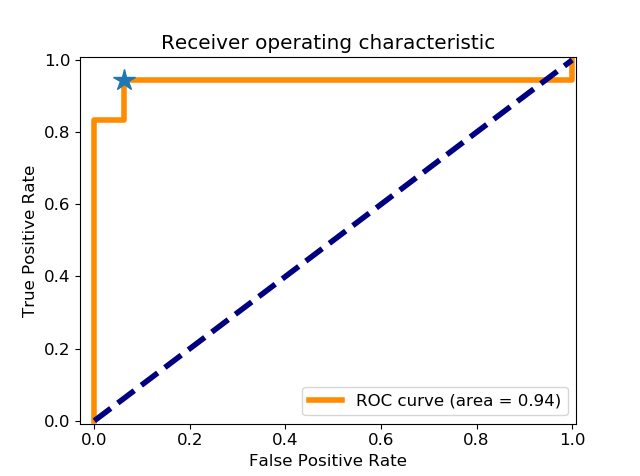}
	\caption[The ROC curve for the diagnosis of CAD on a per-patient level]{The ROC analysis for the diagnosis of CAD on a per-patient level.}
	\label{fig:ROC_global}
\end{figure}

\begin{figure}[h!]
	\centering
	\includegraphics[width=.9\textwidth]{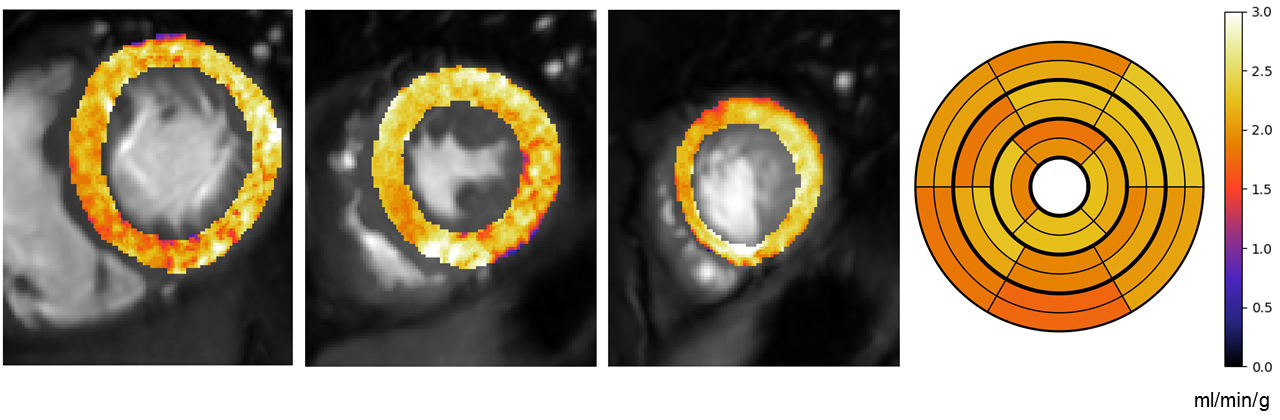}
	\caption[Quantitative MBF maps and 32 segment representation for a CAD- patient.]{Quantitative MBF maps and 32 segment representation for a CAD- patient.}
	\label{fig:cnt}
\end{figure}

\begin{figure}[h!]
	\centering
	\includegraphics[width=.9\textwidth]{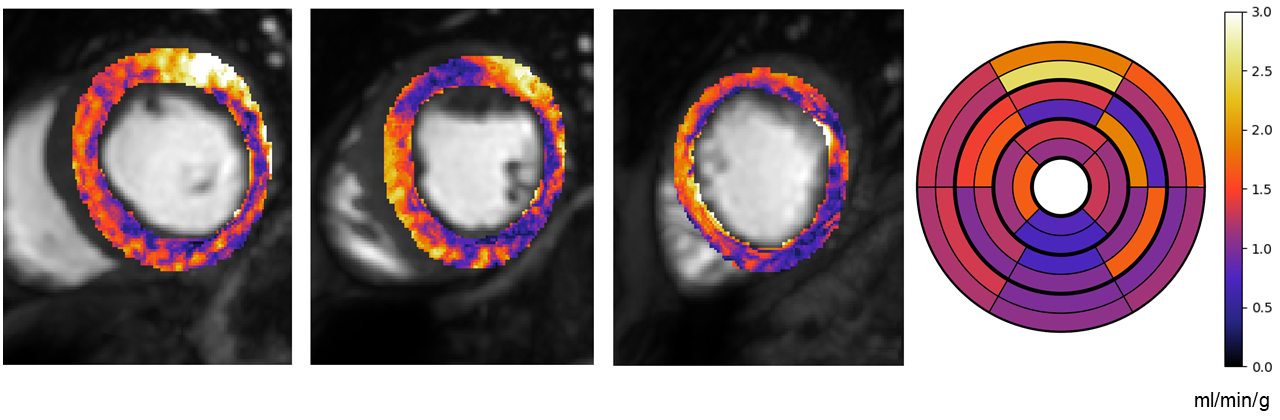}
	\caption[Quantitative MBF maps and 32 segment representation for a CAD+ patient.]{Quantitative MBF maps and 32 segment representation for a CAD+ patient.}
	\label{fig:3vd}
\end{figure}

In the setting of patients with known or suspected CAD, the ROC curve in Figure~\ref{fig:ROC_global} shows that quantitative stress perfusion CMR distinguishes well between CAD- and CAD+ patients. The area under the curve (AUC) of the ROC curve was 0.94 and using an threshold of 1.34 \si{\milli\litre\per\min\per\gram} to detect CAD achieved a sensitivity of 94.4\%, a specificity of 93.8\%, and a diagnostic accuracy of 91\%.
Figures~\ref{fig:cnt} and~\ref{fig:3vd} show examples of quantitative perfusion maps and the corresponding segmental analysis for patients correctly classified as CAD- and CAD+, respectively.

\begin{figure}[h!]
	\centering
	\includegraphics[width=.9\textwidth]{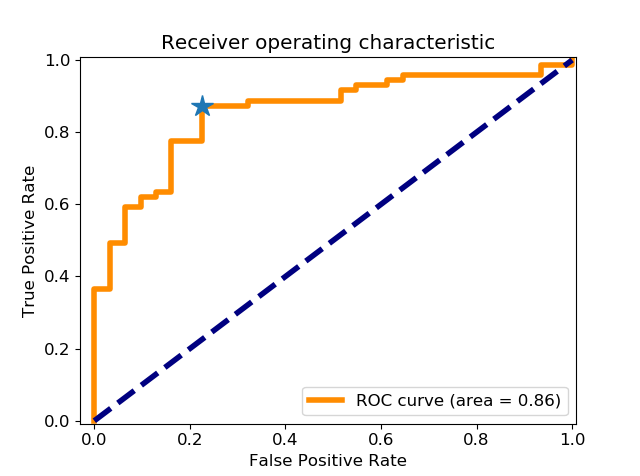}
	\caption[The ROC curve for the diagnosis of CAD on a per-vessel level]{The ROC analysis for the diagnosis of CAD on a per-vessel level.}
	\label{fig:ROC_vessel}
\end{figure}

\begin{figure}
	\centering
	\includegraphics[width=.9\textwidth]{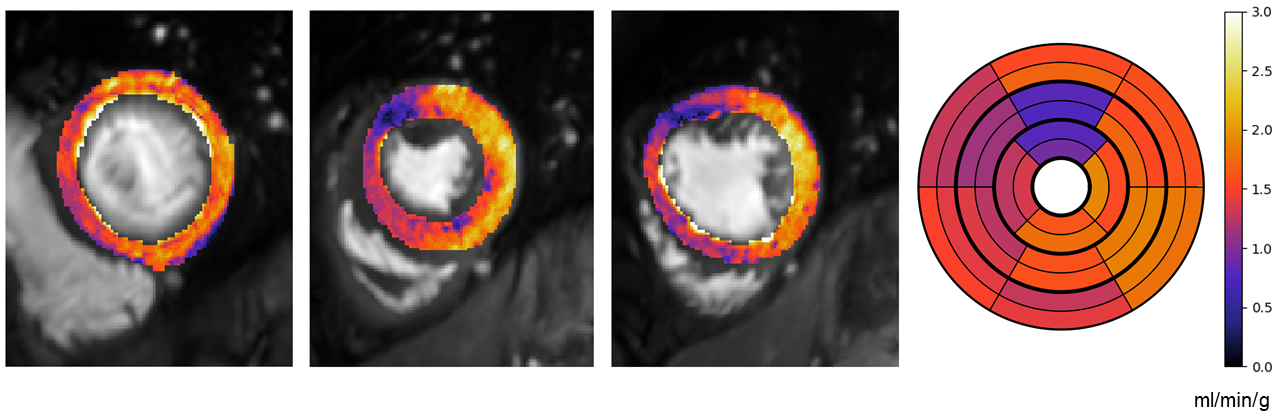}
	\caption[An example of a misclassified patient]{An example of a misclassified patient. The patient has triple vessel disease and while they are correctly identified as CAD+ by the MBF values, they are misclassified as having single vessel disease (LAD). There is significant lesions but not significant ischaemia in the LCx and RCA.}
	\label{fig:3vd_miss}
\end{figure}

This diagnostic power is reduced on the per-vessel assessment. The ROC curve is shown in Figure~\ref{fig:ROC_vessel} and the AUC of the ROC curve is 0.86, with the threshold of 1.31 \si{\milli\litre\per\min\per\gram} giving a sensitivity of 87.3\%, a specificity of 77.1\%, and a diagnostic accuracy of 83.3\%.
The mean MBF in regions supplied by vessels with a stenosis was $1.19 \pm 0.28$ \si{\milli\litre\per\min\per\gram}. This was significantly higher in regions supplied by vessels without a stenosis at $1.8 \pm 0.59$ \si{\milli\litre\per\min\per\gram} ($p < 0.01$). 
An example patient with misclassified vessels is shown in Figure~\ref{fig:3vd_miss}. This patient has triple vessel disease but is classified as being ischaemic in only the LAD territory. The ischaemia in the circumflex region is subtle and restricted to sub-endocardial region in the mid slice which is not picked up by the segmental analysis. Furthermore, the ischaemia in the RCA territory is visible in the septal/inferoseptal segments of the mid and apical slices but the perfusion values are in an intermediate range and do not fall below the threshold for ischaemia.
\begin{figure}
	\centering
	\includegraphics[width=.9\textwidth]{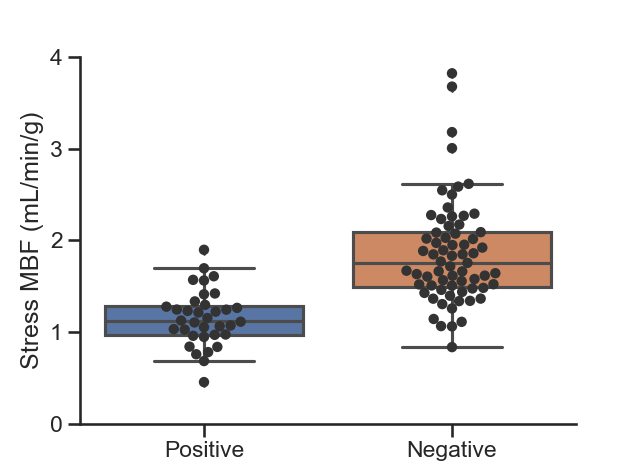}
	\caption[The distribution of MBF values recorded for each vessel in the CAD+ patients.]{The distribution of MBF values recorded for each vessel in the CAD+ patients, divided between positive (for CAD) and negative.}
	\label{fig:pos_neg}
\end{figure}

\begin{figure}
	\centering
	\includegraphics[width=.9\textwidth]{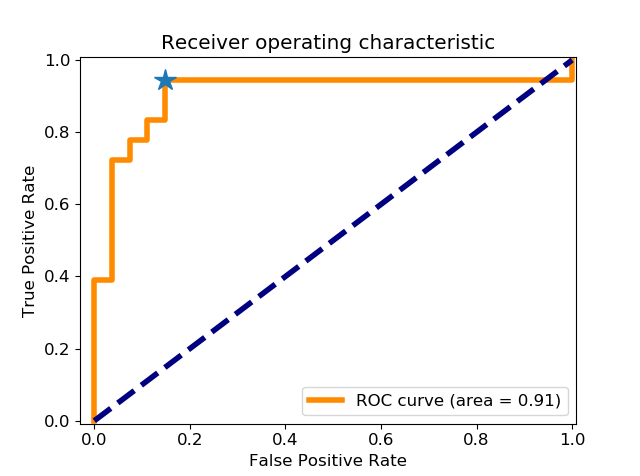}
	\caption[The ROC analysis for the detection of ischaemia on a per-patient level.]{The ROC analysis for the detection of ischaemia on a per-patient level.}
	\label{fig:ROC_isch}
\end{figure}

With the inclusion of patients with MVD, the ability of quantitative stress perfusion CMR to detect patients with myocardial ischaemia is good, as demonstrated by the ROC curve in Figure~\ref{fig:ROC_isch}. The AUC of the ROC analysis was 0.91. The optimal threshold for detecting ischaemia was 1.34 \si{\milli\litre\per\min\per\gram}. This threshold gave a sensitivity of 94.4\%, a specificity of 85.2\% and an overall diagnostic accuracy of 86.7\%.

\begin{figure}
	\centering
	\includegraphics[width=.9\textwidth]{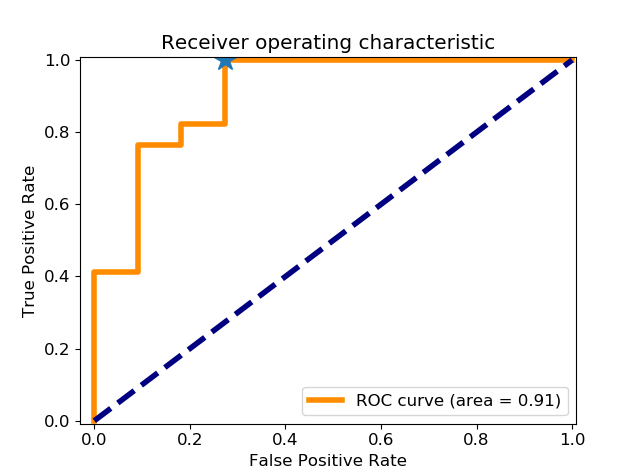}
	\caption[The ROC analysis for the detection of MVD on a per-patient level.]{The ROC analysis for the detection of MVD on a per-patient level.}
	\label{fig:ROC_mvd}
\end{figure}

\begin{figure}
	\centering
	\includegraphics[width=.9\textwidth]{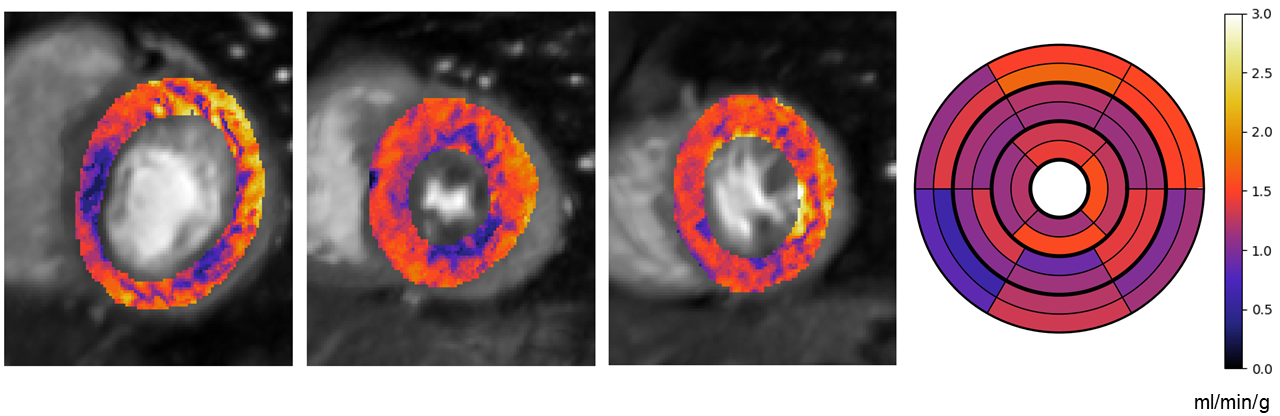}
	\caption[Quantitative MBF maps and 32 segment representation for a MVD patient.]{Quantitative MBF maps and 32 segment representation for a MVD patient.}
	\label{fig:mvd1}
\end{figure}

\begin{figure}
	\centering
	\includegraphics[width=.9\textwidth]{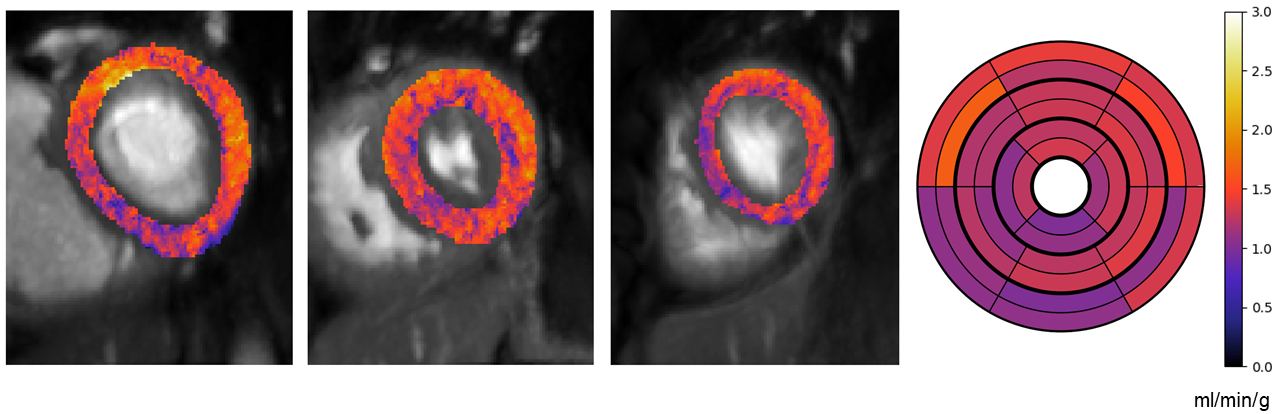}
	\caption[Quantitative MBF maps and 32 segment representation for a MVD patient.]{Quantitative MBF maps and 32 segment representation for a MVD patient.}
	\label{fig:mvd2}
\end{figure}

If only the patients with anatomically smooth coronary vessels are included, it gives a realistic clinical setting to test if quantitative stress CMR can distinguish between CAD- and MVD patients. The ROC analysis for this situation is shown in Figure~\ref{fig:ROC_mvd}. The AUC is 0.91, sensitivity is 100\% and specificity of 71.7\% so that the overall diagnostic accuracy is 85.7\%.
In the patients with MVD, the typical diffuse, sub-endocardial ischaemia is seen, particularly in the (systolic) mid-slice \cite{Panting2002}. Examples of these patients are shown in Figures~\ref{fig:mvd1} and~\ref{fig:mvd2}. 

\section{Discussion}
This is a small proof of principle study to establish the feasibility of fully-automatic quantitative stress perfusion CMR using the methods presented in Chapters 4, 5, and 6. That is, it has been shown that a stress perfusion CMR image series can be automatically processed, motion compensated, segmented, and tracer-kinetic models can be fit to automatically infer myocardial blood flow which has a high level of diagnostic accuracy for detecting CAD and MVD. It removes the subjectivity and laborious processing of the analysis. If these results are found to generalise well to larger and independent patient groups, this will really make the case for the widespread clinical adoption of stress perfusion CMR. 

The diagnostic accuracy and even the cut-off threshold values found are very similar to recent, concurrently run, studies \cite{Hsu2018,Knott2019}. The main advantage of this study is that it was the first to be truly automatic, including myocardial segmentation and disease classification.

There are, however, limitations to this work. Most notable are the small sample size and the comparison of the functional perfusion test to an anatomical reference standard. It is well discussed that the anatomical measurements do not always correlate well to their functional significance. Therefore, a follow-up study is warranted in which quantitative stress CMR is compared to appropriate reference standards: fractional and coronary flow reserve. 

\begin{figure}[htbp!]
	\centering
	\includegraphics[width=.9\textwidth]{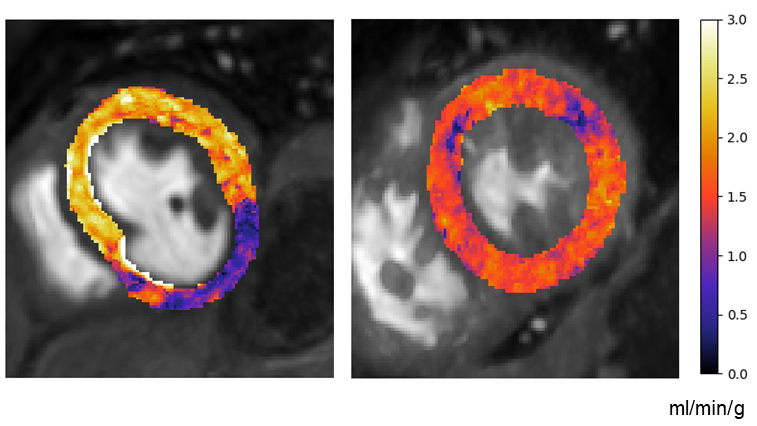}
	\caption[Example MBF maps from two patients.]{Example MBF maps from two patients. The patient on the left has extensive ischaemia arising from disease in one vessel. The patient on the left has two vessel disease but much less ischaemia. This did not reach the derived threshold for CAD and is misclassified.}
	\label{fig:compare}
\end{figure}

There is also further future work required. This study was conducted with dual-bolus acquisitions, methods for dual-sequence acquisitions are now becoming available and should ease the workflow of future studies \cite{Sanchez-Gonzalez2015} but the different acquisition may need different cut-off thresholds. The diagnostic accuracy could also benefit from future developments. While the per-patient diagnostic accuracy is very high, this drops significantly when assessing which vessels are diseased. A physiological cause of this may be some microvascular dysfunction causing ischaemia in the remote territories of patients with CAD leading to false positives. It is also likely a consequence of the limitations of the segmental analysis. Small areas of ischaemia may be averaged out making a segment to be a false negative. False positives are also possible when the area of ischaemia extends into the neighbouring territory. Two examples of patients with these characteristics are shown in Figure~\ref{fig:compare}. The patient on the left has a single RCA lesion and the patient on the right has LCx and LAD lesions. Despite the patient on the left having disease in fewer vessels, they have more ischaemia and the amount of disease is over-estimated as the ischaemia extends into the LCx territory. On the right, the disease is not picked up as the ischaemic burden is not high enough. However, the diagnosis would be clear to a visual reader based on this quantitative maps and the use of machine learning classifiers to mimic this may lead to more refined diagnoses. The ability detect CAD from CAD- patients and MVD from CAD- is shown but the ultimate aim would be to be able to distinguish the three groups from each other and this may benefit from a similar machine learning classifier.

\chapter{Conclusion}

\section{Thesis summary}
This thesis has dealt with the challenging problem of myocardial perfusion quantification from stress perfusion CMR images. Myocardial perfusion CMR is a powerful diagnostic tool for coronary artery disease and has the potential to play a big role in the management of patients with microvascular dysfunction. It has been, so far, limited by the difficulty of interpreting the images (as well as factors such as the lack of availability, and the problem of reimbursement in the United States). Quantitative stress perfusion CMR is potentially the solution to this as it is not subjective and thus requires less expertise. 

This idea is not new and perfusion quantification has been an active research topic for many years \cite{Wilke1997}. A major drawback of all these early solutions, however, was that they required a lot of manual processing, such as segmenting the AIF, segmenting the myocardium, and motion compensation. Biglands et al. \cite{Biglands2018} reported that it took roughly one hour per patient to manually correct the myocardial contours for motion. This level of manual interaction is clearly not feasible on a routine basis in the clinic. Furthermore, the reliability of the analysis, particularly the model fitting \cite{Buckley2002,Likhite2017}, has been often questioned. This motivated the aim of this thesis, which was to make the analysis fast, automatic, and more reliable. This work has led to a solution for quantitative stress perfusion CMR that is robust and truly automated.

Chapter 4 presented an approach for the compensation of respiratory motion which often precludes accurate quantification. Motion compensation is particularly challenging in DCE-MRI as it uses saturation recovery sequences and injections of contrast boluses. The contrast bolus passes rapidly through the region of interest leaving a series of time dynamics with very different levels and regions of contrast enhancement. As traditional image registration approaches assume there is a constant mapping between anatomical regions and signal intensity values, this assumption is invalidated by the passing contrast. Outside of where and when the contrast is passing, the signal is saturated. With little signal to work with, intensity-based registrations are sensitive to noise and subject to failing.  

The data decomposition technique robust PCA is shown to be able to separate the dynamic contrast-enhancement from the baseline anatomical signal. The rigid motion can then be estimated in the absence of the dynamic contrast-enhancement and subsequently applied to the original image series. The remaining motion appears to be noise-like and can be removed using a PCA-decomposition and the image series can again be registered to the de-noised motionless synthetic image series to remove any remaining motion. It is well documented that the validation of image registrations is challenging \cite{Pluim2017}, in this work the quantitative evaluation is based on assumptions about motionless image series. It is shown that the time-intensity curves evolve more smoothly and that the rest perfusion maps show more spatial homogeneity after motion compensation. Perhaps more telling is that expert readers visually scored the quality of the motion compensation higher than the state-of-the-art approach.

In Chapter 5, the first deep learning-based pipeline for stress perfusion CMR was described. This leverages the huge advances made in computer vision and image processing, based on convolutional neural networks, to give a previously unseen level of autonomy to the processing. The pipeline seeks to mimic how a manual operator would have processed the data in the past. A first CNN picks the time dynamic with maximum contrast, this image is then fed to a further network which computes a bounding box around the LV and myocardium. Motion compensation is then performed within this bounding box. The resulting image series is finally segmented and the RV insertion point is detected. The training labels were produced by expert operators with a vast amount of experience in perfusion quantification. The algorithms were validated against labels from the same operators on the held-out test set. The automated deep learning approach was shown to closely match the expert labels at all stages. Most importantly, there was also a good agreement between the derived MBF values from both the automated and manual systems.

Chapter 6 aimed to improve the reliability of the kinetic parameters being inferred. This was done by enforcing well-known and physiological prior information. In particular, the knowledge that neighbouring pixels are likely to have similar kinetics was exploited. This prior information can be naturally included in a Bayesian inference framework and was combined with probabilistic constraints to ensure the estimated parameters were within physiological ranges. Since ground-truth kinetic values are not available \textit{in vivo}, this was first validated in a series of simulated studies and shown to outperform the traditional non-linear least squares fitting estimates. Visually, the parameters estimated with Bayesian inference in patients outperformed the least-squares parameters and in a small cohort they were shown to match the clinical assessment of the patients.

The main validation for the combined approaches from Chapters 4, 5, and 6 comes in Chapter 7. Here it is shown, in a cohort of 47 patients that the automatically derived quantitative MBF values can diagnose CAD and in the setting of patient with smooth coronary arteries, it can diagnose coronary MVD. As discussed, this study is somewhat limited and a follow-up study in a larger cohort versus a functional marker, such as FFR, is warranted.

\section{Context}
Stress perfusion CMR has undoubted high diagnostic accuracy \cite{Greenwood2012} and has outperformed SPECT in randomised trials \cite{Schwitter2013a}. It has been shown to have high prognostic value and be cost-effective \cite{Kwong2019}. Perhaps most importantly, the results of MR-INFORM \cite{Nagel2019} showed that it is non-inferior to invasive FFR measurements for the management of patients \cite{Nagel2019}. Additional advantages over the alternative nuclear perfusion imaging techniques are the higher spatial resolution and lack of ionising radiation. Thus, it should play a significant role in the management of patients with suspected CAD. The use of stress perfusion CMR should only benefit from the analysis being made easier and less subjective.

This works comes at an important time. Through-out the last 20 years there has been a trend away from anatomical imaging towards functional assessments of CAD. This trend made sense based on the intuition that there must be demonstrable ischaemia for a patient to benefit from being revascularised. It also seemed to be backed up by the FAME I and FAME II trials which showed better outcomes for patients guided by functional or ischaemia tests \cite{Pijls2010,DeBruyne2012}.

However, recently, anatomical imaging has undergone a resurgence based on the results of the PROMISE trial \cite{Douglas2015} (and followed-up by the SCOT-HEART trial \cite{Williams2017}) which showed that patients can be equally well managed by anatomical CCTA imaging. This resurgence has been accelerated by the (mis-)interpretation of the results of the ISCHEMIA trial \cite{Maron2020}. Despite the fact that the trial was designed to compare invasive interventions versus optimal medical therapy and not to compare imaging tests, it was reported that there was no correlation between ischaemia and benefit from revascularisation. It is important to observe that the ISCHEMIA  used local reads for the stress tests. As discussed extensively in this thesis, the diagnostic accuracy of stress imaging tests falls dramatically outside of highly experienced centres and this is likely to have impacted the results. This is clearly an area where emerging technology, such as quantitative perfusion CMR, has the potential to add real value. Furthermore, the use of CCTA in ISCHEMIA to exclude left main disease has led to it being marketed as the first in line imaging test for CAD. This is all in spite of CCTA only being able to reliably rule out CAD and not having a positive predictive value, as it does not infer the presence of ischaemia.

Finally, a further major advantage of ischaemia testing over anatomical imaging is in patients that present with anatomically smooth coronary arteries and angina, from microvascular dysfunction. MVD causes ischaemia \cite{Rahman2019} and as was shown in this thesis, there is a role for ischaemia imaging in these patients. The quantitative analysis may finally allow these patients to be managed on the basis of a non-invasive test.

\section{Future work}

The methods presented in this thesis represent a significant improvement on methods previously used in the field. The robust motion compensation alleviates one of the significant challenges and automated processing makes it more accessible. This is the first deep learning system for stress perfusion CMR, though more are now becoming available \cite{Xue2020a}. However, there are still opportunities for future work in the field. As discussed, the output of the quantification would be a suitable input for a further machine learning classifier for disease classification and staging. This would circumnavigate the discussed limitations of the AHA segment model. A limitation of the proposed methods is the processing time taken for the analysis, the image registration and Bayesian inference are computationally intensive. These steps could be replaced with deep learning systems \cite{scannell2019deep} which are quicker to deploy on new data. 

The power of deep learning could be utilised in a range of other tasks from which the analysis would benefit. This includes the image reconstruction, to allow images to be acquired faster, and super resolution algorithms to increase both the in-plane spatial resolution and the LV coverage. There has also been a recent interest in developing quality control measures for automated processing systems \cite{Robinson2017,Ruijsink2020}. The system presented in this thesis is not infallible and would benefit from an assessment of image quality prior to processing and a method to detect failed cases.

In conclusion, while this work has laid the foundations of fully automatic quantitative stress perfusion CMR, much more work can be done to make the analysis more accurate, reliable, and fast. This further work will likely leverage the tremendous improvements in image processing made possible by advances in machine learning. The results presented, and in the literature, are promising, but in the end trials will be needed to validate it, preferably including less specialised centres.


\begin{spacing}{1.}


\bibliographystyle{ieeetr}
\cleardoublepage
\bibliography{References/references,References/library} 



\end{spacing}


\begin{appendices} 

\end{appendices}

\printthesisindex 

\end{document}